\documentclass[preprints,article,accept,oneauthor]{mdpi} 
\pdfoutput=1

\firstpage{1} 

\pubvolume{vv}
\issuenum{X}
\articlenumber{Y}
\pubyear{2020}
\copyrightyear{2020}
\history{}

\Title{Polynomial Representations of High-Dimensional Observations of Random
  Processes}

\Author{Pavel Loskot$^{1}$*\orcidA{}}

\AuthorNames{Pavel Loskot}

\address{$^{1}$ \quad ZJU-UIUC, Haining, China; loskot@gmail.com}
\corres{Correspondence: loskot@gmail.com}

\newcommand{\eref}[1]{(\ref{#1})} 
\newcommand{\fref}[1]{Figure~\ref{#1}} 
\newcommand{\pref}[1]{p.~\pageref{#1}} 

\newcounter{conjecture}
\newtheorem{Conjecture}[conjecture]{Conjecture}

\usepackage{psfrag}
\usepackage{amssymb,amsfonts,mathrsfs,mathptm,bm,mathtools}

\newcommand{\R}{\mathcal{R}} 
\newcommand{\Rp}{\mathcal{R}^{\!+}} 
\newcommand{\N}{\mathbb{N}} 

\newcommand{\eee}{\,{\operatorname{e}}} 
\DeclareMathOperator*{\argmin}{argmin} 
\DeclareMathOperator*{\argmax}{argmax} 
\newcommand{\jj}{{\operatorname{j}}} 
\newcommand{\df}{{\,\operatorname{d}}} 

\newcommand{\br}{\begin{array}}
\newcommand{\er}{\end{array}}
\newcommand{\eqtext}[1]{\overset{\mbox{\footnotesize #1}}{=}} 

\newcommand{\Prob}[1]{\operatorname{Pr}\!\left(#1\right)} 
\newcommand{\E}[1]{{\operatorname{E}}\!\left[#1\right]} 
\newcommand{\var}[1]{{\operatorname{var}}\!\left[#1\right]} 
\newcommand{\cov}[1]{{\operatorname{cov}}\!\left[#1\right]} 
\newcommand{\corr}[1]{{\operatorname{corr}}\!\left[#1\right]} 

\newcommand{\norm}[1]{\left\lVert#1\right\rVert}

\newcommand{\bY}{\bar{Y}}
\newcommand{\bX}{\bar{X}}
\newcommand{\vX}{\bm{X}}
\newcommand{\vXb}{\bar{\vX}}
\newcommand{\vC}{\bm{C}}
\newcommand{\vT}{\bm{T}}
\newcommand{\vU}{\bm{U}}
\newcommand{\Rb}{\bar{R}}
\newcommand{\vP}{\bm{P}}
\newcommand{\Ph}{\hat{P}}
\newcommand{\Pb}{\bar{P}}
\newcommand{\vPh}{\hat{\vP}}
\newcommand{\vw}{\bm{w}}
\newcommand{\wb}{\bar{w}}
\newcommand{\Xb}{\bar{X}}
\newcommand{\Wb}{\bar{W}}
\newcommand{\vK}{\bm{K}}
\newcommand{\vZs}{\bm{0}}
\newcommand{\vOs}{\bm{1}}
\newcommand{\vI}{\bm{I}}
\newcommand{\vu}{\bm{u}}
\newcommand{\vx}{\bm{x}}
\newcommand{\vxb}{\bar{\bm{x}}}
\newcommand{\vQ}{\bm{Q}}
\newcommand{\vn}{\bm{n}}
\newcommand{\vm}{\bm{m}}
\newcommand{\vh}{\bm{h}}
\newcommand{\vt}{\bm{t}}
\newcommand{\Yb}{\bar{Y}}
\newcommand{\vs}{\bm{s}}
\newcommand{\vLam}{\bm{\Lambda}}

\newcommand{\conv}{\circledast}
\newcommand{\MPP}{\mathrm{2MP}}
\newcommand{\MP}{\mathrm{1MP}}
\newcommand{\MA}{\mathrm{MA}}
\newcommand{\Scos}{S_{\mathrm{cos}}}
\newcommand{\Smnk}{S_{\mathrm{mnk}}}
\newcommand{\mScos}{\bar{S}_{\mathrm{cos}}}
\newcommand{\mSmnk}{\bar{S}_{\mathrm{mnk}}}
\newcommand{\tSmnk}{\tilde{S}_{\mathrm{mnk}}}
\newcommand{\bmu}{\bar{\mu}}
\newcommand{\dfM}{\dot{M}}
\newcommand{\II}{\mathcal{I}}
\newcommand{\LSopt}{LS_\mathrm{{opt}}}
\newcommand{\LSapr}{LS_\mathrm{{apr}}}
\newcommand{\Gapr}{G_\mathrm{apr}}
\newcommand{\bl}{b_{\mathrm{l}}}
\newcommand{\bu}{b_{\mathrm{u}}}

\newcommand{\bT}{\bar{T}}
\newcommand{\Sopt}{S_\mathrm{{opt}}}
\newcommand{\Sapr}{S_\mathrm{{apr}}}
\newcommand{\cF}{\mathcal{F}}
\newcommand{\cC}{\mathcal{C}}
\newcommand{\vtau}{\bm{\tau}}
\newcommand{\mse}{\mathrm{MSE}}
\newcommand{\ft}{\tilde{f}}
\newcommand{\tv}{\tilde{v}}
\newcommand{\nx}{n_\mathrm{x}}

\abstract{The paper investigates the problem of performing correlation analysis
  when the number of observations is very large. In such a case, it is often
  necessary to combine the random observations to achieve dimensionality
  reduction of the problem. A novel class of statistical measures is obtained
  by approximating the Taylor expansion of a general multivariate scalar
  function by a univariate polynomial in the variable given as a simple sum of
  the original random variables. The mean value of the polynomial is then a
  weighted sum of statistical central sum-moments with the weights being
  application dependent. Computing the sum-moments is computationally efficient
  and amenable to mathematical analysis, provided that the distribution of the
  sum of random variables can be obtained. Among several auxiliary results also
  obtained, the first order sum-moments corresponding to sample means are used
  to reduce the numerical complexity of linear regression by partitioning the
  data into disjoint subsets. Illustrative examples are provided assuming the
  first and the second order Markov processes.}

\keyword{Least squares; linear regression; Markov process; moment method;
  multivariate function; Taylor expansion}

\begin{document}

\section{Introduction}

The interest in developing and improving statistical methods and models is
driven by the ever increasing volumes and variety of data. Making sense of data
often requires to uncover the existing patterns and relationships in data. One
of the most commonly used universal tools for data exploration is evaluation of
cross-correlation among different data sets and of auto-correlation of data
sequence with itself. For example, correlation values can be utilized to carry
out sensitivity analysis, to forecast future values, to visualize complex
relationships among subsystems, and to evaluate other important spatio-temporal
statistical properties. However, it has been well established that correlations
do not imply causalization.

The correlations measure linear dependencies between pairs of random variables
which is implicitly exploited in linear regression. The correlation values rely
on estimated or empirical statistical moments such as sample mean and sample
variance. If the mean values are removed from data, the correlations are
referred to as covariances. The pairwise correlations can be represented as a
fully connected graph with edges parameterized by time shifts, and possibly by
amplitude adjustments between the corresponding data sequences. However, the
graph may become excessively complex when the number of random variables
considered is large; for example, when analyzing multiple long sequences of
random observations.

The problem of extending the notion of correlations and covariances to more
than two random variables has been considered previously in literature. In
particular, the multirelation has been defined as a measure of linearity for
multiple random variables in \cite{Drezner95}. This measure is based on a
geometric analysis of linear regression. Under the assumption of multivariate
Student-t distribution, the statistical significance of the multirelation
coefficient is defined in terms of eigenvalues of the correlation matrix in
\cite{Dear97}. An univariate correlation measure for multiple random variables
is defined in \cite{Geis96} to be a sum of elements on the main diagonal of the
covariance matrix. Linear regression is again utilized in \cite{Abdi07} to
define the multiple correlation coefficient. It is derived from the coefficient
of determination of linear regression as a proportion of the variance in the
dependent variable which is predictable from the independent variables.

The distance correlation measure between two random vectors based on the
hypothesis testing of independence of random variables is proposed in
\cite{Szekely07}. Different operations on random variables including ordering,
weighting and a non-linear transformation are assumed in \cite{Merigo11} to
define a family of Minkowski distance measures for random vectors. A maximal
correlation measure for random vectors is introduced in \cite{Nguyen14}. It
generalizes the concept of maximal information coefficient while non-linear
transformations are also used to allow assessment of non-linear correlations.
Similarly to \cite{Geis96}, the sample cross-correlation between multiple
realizations of two random vectors is shown in \cite{Josse16} to be
proportional to the sum of diagonal elements of the product of the
corresponding correlation matrices. The reference \cite{Shu16} investigates
two-dimensional general and central moments for random matrices which are
invariant to similarity transformations. Two new distance measures for random
vectors are defined in \cite{Bottcher19} assuming the joint characteristic
function. These measures can be also used to test for statistical independence
of the random vectors. The most recent paper \cite{Wang20} derives a linear
correlation measure for multiple random variables using determinants of the
correlation sub-matrices.

This brief literature survey indicates that there are still no commonly
accepted correlation measures for multiple random variables. The measures
considered in the literature are either rigorously derived but mathematically
rather complicated, or there are many modifications of the existing, simpler,
but well understood measures. In this paper, it is shown that by constraining
the complexity of multivariate Taylor series to reduce the number of its
parameters or degrees-of-freedom, the Taylor series can be rewritten as a
finite degree univariate polynomial. The independent variable of the polynomial
is equal to a simple sum of the random variables considered. The polynomial
defines a many-to-one transformation of multiple random variables to another
scalar random variable. The polynomial coefficients are real-valued constants,
and they are application dependent. The mean value of the polynomial represents
a broad class of polynomial measures which can be used for any number of random
variables. The mean value of each polynomial element corresponds to a general
or central moment of the sum of random variables. Therefore, these moments are
referred to here as sum-moments. In case of multiple random vectors, similarly
to computing the correlation or covariance matrix by first concatenating the
vectors into one long vector, the sum-moments can be readily defined and
computed for such a concatenated random vector. The main advantages of assuming
sum-moments to study statistical properties of multiple random variables or
multiple random vectors are the clarity in understanding their statistical
significance, mathematical simplicity of their definitions, and as long as the
distribution of the sum of random variables can be found, the closed-form
expression for the sum-moments can be obtained.

Before introducing polynomial representations of random vectors in Section 4
together with central and general sum-moments, a number of auxiliary results
are presented in Section 2 and Section 3. In particular, Section 2 summarizes
the key results and concepts from the literature concerning stationary random
processes, their parameter estimation via linear regression and method of
moments, and how to generate the 1st and the 2nd order Markov processes is also
discussed. Section 3 extends the results from Section 2 by deriving additional
results which are used in Section 4 or in Section 5 such as a low complexity
approximation of linear regression and a procedure to generate multiple
Gaussian processes with defined auto-correlation and cross-correlation. The
main results of the paper are obtained in Section 4 including defining a class
of polynomial statistical measures and sum-moments for multiple random
variables and random processes. Other related concepts involving sums of random
variables are also reviewed. Section 5 provides several examples to illustrate
and evaluate the results obtained in the paper. The paper is concluded in
Section 6.

\section{Background}

This section reviews key concepts and results from the literature which are
used to develop new results in the subsequent sections. Specifically, the
following concepts are briefly summarized: stationarity of random processes,
estimation of general and central moments and of correlation and covariance
using the method of moments, definition of cosine similarity and Minkowski
distance, parameter estimation via linear regression, generation of the 1st and
the 2nd order Markov processes, and selected properties of polynomials and
multivariate functions are given. Moreover, note that more straightforward
proofs for some lemmas are only indicated, and not fully and rigorously
elaborated.

\subsection{Random processes}

Consider a real-valued one-dimensional random process $x(t)\in\R$ over a
continuous time $t\in\R$. The process is observed at $N$ discrete time
instances $t_1<t_2<\ldots<t_N$ corresponding to $N$ random variables
$X_i=x(t_i)$, $i=1,\ldots,N$. The random variables
$\vX=\{X_1,\ldots,X_N\}\in\R^N$ are completely statistically described by their
joint density function $f_x(\vX)$, so that $f_x(\vX)\geq 0$ and
$\int_{\R^N} f_x(\vX)\df\vX=1$. The process $x(t)$ is further assumed to be
$K$-th order stationary \cite{Gardner90}.
\begin{Definition}
  A random process $x(t)$ is $K$-th order stationary, if,
  \begin{equation*}
    f_x(X_1,\ldots,X_K;t_1,\ldots,t_K)=f_x(X_1,\ldots,X_K;t_1+t_0,\ldots,t_K+t_0)
    \quad \forall t_0\in\R.
  \end{equation*}
\end{Definition}
\begin{Lemma}
  The $K$-th order stationary process is also $(K-k)$-th order stationary,
  $k=1,2,\ldots,K-1$, for any subset
  $\{X_{i_1},\ldots,X_{i_{K-k}}\}\subseteq \{X_1,\ldots,X_N\}$.
\end{Lemma}
\begin{proof}
  The unwanted random variables can be integrated out from the joint density.
\end{proof}
Unless otherwise stated, all observations of random processes are assumed to be
stationary.

The expectation, $\E{X}=\int_\R x f_x(x)\df x$, of a random variable $X$ is a
measure of its mean value. A linear correlation between between two random
variables $x(t_1)$ and $x(t_2)$ is defined as \cite{Gardner90},
\begin{equation*}
  R_x(t_1,t_2)= \corr{x(t_1) x(t_2)} = R_x(t_2-t_1)=R_x(\tau).
\end{equation*}
The (auto-) covariance measures a linear dependency between two zero-mean
random variables, i.e.,
\begin{equation*}
  C_x(t_1,t_2)= \cov{x(t_1),x(t_2)}= \E{\left(x(t_1)-\E{x(t_1)}\right)
    \left(x(t_2)-\E{x(t_2)}\right)} = C_x(t_2-t_1).
\end{equation*}
It can be shown that the maximum of $C_x(\tau)$, $\tau=t_2-t_1$, occurs for
$\tau=0$, corresponding to the variance of a stationary process $x(t)$, i.e.,
\begin{equation*}
  C_x(0)=\var{x(t)}=\E{\left(x(t)-\E{x(t)}\right)^2}\quad \forall t\in\R.
\end{equation*}
The covariance $C_x(\tau)$ can be normalized, so that,
$-1\leq C_x(\tau)/C_x(0)\leq 1$. Furthermore, for real-valued regular
processes, the covariance has an even symmetry, i.e., $C_x(\tau)=C_x(-\tau)$,
\cite{Gardner90}.

For $N>2$, it is convenient to define a random vector,
$\vX= [X_1,\ldots,X_N]^T$, and its mean as, $\vXb=\E{\vX}$ where $(\cdot)^T$
denotes the vector or matrix transpose. The corresponding
covariance matrix,
\begin{equation*}
  \vC_x=\E{(\vX-\vXb)(\vX-\vXb)^T}\subseteq\R^{N\times N}
\end{equation*}
has as its elements the pairwise covariances, $[\vC_x]_{ij} = C_x(t_j-t_i)$,
$i,j=1,2,\ldots,N$.

Assuming now the case of two $K$-th order stationary random processes $x_1(t)$
and $x_2(t)$, and the discrete time observations,
\begin{equation*}
  \begin{split}
    X_{1i} &= x(t_{1i}),\quad i=1,2,\ldots,N_1 \\
    X_{2i} &= x(t_{2i}),\quad i=1,2,\ldots,N_2.
  \end{split}
\end{equation*}
Using the set notation, $\{X_i\}_{1:K}=\{X_1,X_2,\ldots,X_K\}$, the jointly or
mutually stationary processes imply time-shift invariance of their joint
density function.
\begin{Definition}
  Random processes $x_1(t)$ and $x_2(t)$ are $K$-th order jointly stationary,
  if,
  \begin{equation*}
    f_x(\{X_{1i}\}_{1:K},\{X_{2i}\}_{1:K};\{t_{1i}\}_{1:K},
    \{t_{2i}\}_{1:K}) = f_x(\{X_{1i}\}_{1:K},\{X_{2i}\}_{1:K};
    \{t_{1i}+t_0\}_{1:K},\{t_{2i}+t_0\}_{1:K}) \quad \forall t_0\in\R
  \end{equation*}
  is satisfied for all subsets, $\{X_{1i}\}_{1:K}\subseteq \{X_{1i}\}_{1:N_1}$,
  $K\leq N_1$, and $\{X_{2i}\}_{1:K}\subseteq \{X_{2i}\}_{1:N_2}$, $K\leq N_2$.
\end{Definition}
\begin{Lemma}
  The $K$-th order joint stationarity implies the joint stationarity of all
  orders smaller than $K$.
\end{Lemma}
\begin{proof}
  The claim follows from marginalization of the joint density function to
  remove unwanted variables.
\end{proof}

The cross-covariance of random variables $X_1=x_1(t_1)$ and $X_2=x_2(t_2)$
being discrete time observations of the jointly stationary random processes
$x_1(t)$ and $x_2(t)$ is defined as,
\begin{equation*}
  C_{x_1x_2}(t_1,t_2)=\cov{X_1,X_2}= \cov{\left(x(t_1)-\E{x(t_1)}\right)
    \left(x(t_2)-\E{x(t_2)}\right)} = C_{x_1x_2}(t_2-t_1).
\end{equation*}
The cross-covariance can be again normalized, so it is bounded as,
\begin{equation*}
  -1\leq C_{x_1x_2}(\tau)/\sqrt{\var{x_1(t)}\var{x_2(t)}}\leq 1.
\end{equation*}
Note that, unlike auto-covariance, the maximum of $C_{x_1x_2}(\tau)$ can occur
for any value of the argument $\tau$.

The covariance matrix for the random vectors $\vX_1=[X_{1i},\ldots,X_{1N_1}]^T$
and $\vX_2=[X_{2i},\ldots,X_{2N_2}]^T$ having the means, $\vXb_1=\E{\vX_1}$ and
$\vXb_2=\E{\vX_2}$, respectively, is computed as,
\begin{equation*}
  \vC_{x_1 x_2}=\E{(\vX_1-\vXb_1)(\vX_2-\vXb_2)^T}\subseteq\R^{N_1\times N_2}.
\end{equation*}
Its elements are the covariances,
$[\vC_{x_1 x_2}]_{ij}=C_{x_1x_2}(t_{1i}-t_{2j})$, $i=1,\ldots,N_1$,
$j=1,\ldots,N_2$.

In addition to the first order (mean value) and the second order (covariance)
statistical moments, higher order statistics are given by the general and the
central moments, respectively, defined as \cite{Papoulis02},
\begin{equation}\label{eq:q8}
  g_m(X)= \E{|X|^m} \mbox{ and } \mu_m(X)= \E{|X-g_1(X)|^m},\quad m=1,2,\ldots
\end{equation}
where $|\cdot|$ denotes the absolute value, and note, $g_1(X)=\E{X}$. The
positive integer-valued moments \eref{eq:q8} facilitate mathematically
tractable integration, and prevent producing complex numbers, if the argument
is negative. Note also that the absolute value in \eref{eq:q8} is necessary if
$X$ is complex valued, or if $m$ is odd, in order to make the moments to be
real-valued and convex. The central moment can be normalized by the variance
as,
\begin{equation*}
  \mu_m(X)= \E{|X-g_1(X)|^m} / \E{(X-g_1(X))^2}^{m/2}.
\end{equation*}

The cosine similarity between two equal-length vectors,
$\vX_1=[X_{11},\ldots,X_{1N}]^T$ and $\vX_2=[X_{21},\ldots,X_{2N}]^T$, is
defined as \cite{Giller12},
\begin{equation*}
  \Scos(\vX_1,\vX_2) = \frac{\sum_{i=1}^N X_{1i}X_{2i}}
  {\sqrt{\sum_{i=1}^N X^2_{1i}}\sqrt{\sum_{i=1}^N X^2_{2i}}}.
\end{equation*}

The Minkowski distance between two equal-length vectors $\vX_1$ and $\vX_2$ is
defined as the $l_m$-norm \cite{Merigo11}, i.e.,
\begin{equation*}
  \Smnk(\vX_1,\vX_2) = \norm{\vX_1-\vX_2}_m = \Big(\sum_{i=1}^N \left|
    X_{1i}-X_{2i}\Big|^m \right)^{1/m},\quad m=1,2,\ldots
\end{equation*}

\subsection{Estimation methods}

Statistical moments can be empirically estimated from measured data using
sample moments. Such inference strategy is referred to as the method of moments
\cite{Kay93}. In particular, under the ergodicity assumption, the first moment
of random variable $X$ is estimated as a sample mean from the measurements
$X_i=x_i$ , i.e.,
\begin{equation*}
  \E{X} = \lim_{N\to\infty} \frac{1}{N} \sum_{i=1}^N X_i.
\end{equation*}
It is straightforward to show that the sample mean estimator is unbiased and
consistent \cite{Kay93}. More generally, the expected value of a transformed
random variable $h(X)$ is calculated as,
\begin{equation*}
  \E{h(X)} = \lim_{N\to\infty} \frac{1}{N-d} \sum_{i=1}^N h(X_i)
\end{equation*}
where $d$ is the number of degrees of freedom used in the transformation $h$,
i.e., the number of other parameters which must be estimated. For example, the
variance of $X$ is estimated as,
\begin{equation*}
  \var{X} =  \lim_{N\to\infty} \frac{1}{N-1} \sum_{i=1}^N (X-\hat{\bX})^2
\end{equation*}
where $\hat{\bX}$ is the estimated mean value of $X$.

Assuming random sequences $\{X_{1i}\}_{1:N_1}$ and $\{X_{2i}\}_{1:N_2}$, their
auto-covariance and cross-covariance, respectively, are estimated as
\cite{Kay93},
\begin{equation*}
  \begin{split}
    C_x(k) = \E{X_iX_{i+k}} & = \lim_{N\to\infty} \frac{1}{N-k}
      \sum_{i=1}^{N-k} X_i X_{i+k},\quad k\ll N.\\
      C_{x_1x_2}(k) = \E{X_{1i}X_{2(i+k)}} & = \lim_{N\to\infty} \frac{1}{N-k}
      \sum_{i=1}^{N-k} X_{1i} X_{2(i+k)},\quad k\ll N=\min(N_1,N_2).
  \end{split}
\end{equation*}
The condition $k\ll N$ is necessary, since the accuracy of estimates decreases
with $k$.

The parameters of a random process can be estimated by fitting a suitable data
model to the measurements. Denote such data model as, $X_i=M_i(\vP)$,
$i=1,2,\ldots,N$. Assuming the least-squares (LS) criterion, the vector of
parameters, $\vP=[P_1,\ldots,P_D]^T\subseteq\R^D$, is estimated as,
\begin{equation*}
  \vPh = \argmin_{\vP} \sum_{i=1}^N (X_i-M_i(\vP))^2.
\end{equation*}
For continuous parameters, the minimum is obtained using the derivatives, i.e.,
let $\frac{\df}{\df \vP} M_i(\vP)= \dfM_i(\vP)$, and,
\begin{equation*}
  \frac{\df}{\df \vP}  \sum_{i=1}^N (X_i-M_i(\vP))^2 \eqtext{!} 0
  \quad \Leftrightarrow \quad \sum_{i=1}^N \dfM_i(\vPh)\,M_i(\vPh) =
  \sum_{i=1}^N \dfM_i(\vPh)\, X_i. 
\end{equation*}
For a linear data model, $M_i(\vP)= \vw_i^T \vP$ where
$\vw_i=[w_{1i},\ldots, w_{Di}]^T$ are known coefficients, the LS estimate can
be obtained in the closed-form, i.e.,
\begin{equation}\label{eq:z0}
  \vPh = \left( \sum_{i=1}^N \vw_i \vw_i^T \right)^{-1} \sum_{i=1}^N \vw_i X_i
\end{equation}
where $(\cdot)^{-1}$ denotes the matrix inverse.

\subsection{Generating random processes}

The task is to generate a discrete time stationary random process with a given
probability density and a given auto-covariance. The usual strategy is to
generate a correlated Gaussian process followed by a non-linear memoryless
transformation. For instance, the autoregressive (AR) process described by the
second order difference equation with constants $0<a<1$ and $0<b$
\cite{Gardner90}, i.e.,
\begin{equation*}
  x(n) + 2 a\, x(n-1) + a^2\, x(n-2) = b\, u(n)
\end{equation*}
generates the 2nd order Markov process from a zero-mean white (i.e.,
uncorrelated) process $u(n)$. The resulting process $x(n)$ has the
auto-covariance,
\begin{equation}\label{eq:x9}
  C_{\MPP}(k) = \sigma^2 (1-a)^{|k|} (1+a\,|k|) \doteq \sigma^2 \eee^{-a|k|}
  (1+a\,|k|)
\end{equation}
where the variance $\sigma^2$ is set by the coefficient $b$. On the other hand,
the AR process,
\begin{equation*}
  x(n)+a x(n) = b\,u(n)
\end{equation*}
generates the 1st order Markov process with the auto-covariance,
\begin{equation}\label{eq:u1}
  C_{\MP}(k) = \sigma^2 a^{|k|} = \sigma^2 \eee^{-\alpha|k|},\quad
  a=\eee^{-\alpha}.
\end{equation}

\begin{Lemma}\cite{Oppenheim09}\label{lm:3}
  The stationary random process $x(k)$ with auto-covariance $C_x(k)$ is
  transformed by a linear time-invariant system with real-valued impulse
  response $h(k)$ into another stationary random process
  $y(k)= \sum_{i=-\infty}^\infty h(i) x(k-i)\equiv h(k)\conv x(k)$ with
  auto-covariance, $C_y(k)= h(k)\conv h(-k)\conv C_x(k)$. The symbol, $\conv$,
  denotes convolution.
\end{Lemma}
\begin{proof}
  By definition, the output covariance,
  $C_y(k,l)=\E{(y(k)-\E{y(k)})(y(l)-\E{y(l)})}$. Substituting
  $y(k)= \sum_{i=-\infty}^\infty h(i) x(k-i)$, and rearranging,
  $C_y(k-l) = \sum_i \sum_m h(m) C_x(i-m) h(k-i) = h(k)\conv h(-k)\conv
  C_x(k)$.
\end{proof}

\begin{Lemma}
  A stationary random process at the output of a linear or non-linear
  time-invariant system remains stationary.
\end{Lemma}
\begin{proof}
  For any multivariate function $h(\vx)\in\R$ and any $t_0\in\R$, the
  expectation,
  \begin{equation*}
    \E{h(\vx)} = \int_{\R^N} h(\vx) f_X(\vx;t_1,\ldots,t_N)\df \vx =
    \int_{\R^N} h(\vx) f_X(\vx;t_1-t_0,\ldots,t_N-t_0)\df \vx
  \end{equation*}
  assuming Definition 1, and provided that the dimension $N$ is at most equal
  to the stationarity order $K$.
\end{proof}

For shorter sequences, the linear transformation, $\vX=\vT\vU$, can be used to
generate a normally distributed vector $\vX\in\R^N$ having the covariance
matrix, $\vC_x=\vT\vT^T$, from uncorrelated Gaussian vector $\vU\in\R^N$. The
mean, $\E{\vX} = \vT\E{\vU}$. For longer sequences, equivalently, a linear
time-invariant filter can be used as indicated by Lemma 3.

\subsection{Polynomials and multivariate functions}

\begin{Lemma}\label{Boyd04}
  A univariate function $p(x)$ is convex, if and only if,
  $\frac{\df^2}{\df x^2} p(x) = \ddot{p}(x)>0$ for $\forall x\in\R$.
\end{Lemma}
\begin{proof}
  Let $p$ be twice differentiable. The convexity implies,
  $p(\lambda x + (1-\lambda) y)\leq \lambda p(x)+(1-\lambda)p(y)$,
  $0\leq \lambda \leq 1$. Therefore, in the limit as $\lambda\to 0$,
  $p(y)-p(x)\geq \dot{p}(x) (y-x)$, and also, $p(x)-p(y)\geq \dot{p}(y) (x-y)$.
  Therefore, $(\dot{p}(y)-\dot{p}(x))(y-x)\geq 0$. Dividing the last inequality
  by $(y-x)^2>0$, in the limit as $y\to x$,
  $(\dot{p}(y)-\dot{p}(x))/(y-x)\geq 0$, and thus, $\ddot{p}(x)\geq 0$.
  Furthermore, the steps can be reversed to show that $\ddot{p}(x)\geq 0$
  implies convexity.
\end{proof}
Consequently, convex polynomials can be generated as follows.
\begin{Lemma}
  Let $\vQ\in\R^{m\times m}$ be a positive semi-definite matrix, and assume a
  polynomial, $\ddot{p}(x)= \sum_{i=0}^{2m-2} p_i x^i$ for $\forall x\in\R$
  where $p_i = \sum_{k+l=i} \vQ_{kl}$. Then, for any $q_0,q_1\in\R$, the
  polynomial $p(x)$ of degree $2m$,
  \begin{equation*}
    p(x) = q_0 + q_1 x + \sum_{i=0}^{2m-2} \frac{p_i}{(i+1)(i+2)}\, x^{2+i}
  \end{equation*}
  is convex.
\end{Lemma}
\begin{proof}
  Let $\vx=[x^0,x^1,\ldots,x^{m-1}]^T$. Then,
  $\vx^T\vQ\vx= \sum_{i=0}^{2m-2} p_i x^i = \ddot{p}(x)\geq 0$ for
  $\forall \vx$, since $\vQ$ is positive semi-definite Using Lemma 5 concludes
  the proof.
\end{proof}

Assuming a non-negative integer $m\in\{0\}\cup \N_{+}=\{0,1,2,\ldots\}$, define
the following notations to simplify the subsequent mathematical expressions
\cite{Folland10}:
\begin{equation*}
  \begin{array}{ccl}
    \vn &=& \{n_1,n_2,\ldots,n_N\}\in\{\N_{+}\cup 0\}^N \\
    \vx &=& \{x_1,x_2,\ldots,x_N\}\in\R^N \\
    \norm{\vx}_p &=& \left(x_1^p+x_2^p+\cdots+ x_N^p\right)^{1/p}\\
    \norm{\vx}_1 &=& |x_1|+|x_2|+\cdots+|x_N|\\
    |\vx| &=& x_1+x_2+\cdots+x_N \\
    \vh &=& \{h_1,h_2,\ldots,h_N\}\in\R^N \\
    \vh^{\vn} &=& h_1^{n_1}h_2^{n_2}\cdots h_N^{n_N} \\
    \partial^{\vn} f(\vx) &=&
    \partial_1^{n_1}\partial_2^{n_2}\cdots\partial_N^{n_N} f(\vx) =
    \frac{\partial^{|\vn|}}{\partial x_1^{n_1} x_2^{n_2} \cdots x_N^{n_N}}
    f(\vx) \\
    m! &=& \prod_{i=1}^m \, i \\
    \vn! &=& n_1!\,n_2!\cdots n_N!
  \end{array}
\end{equation*}
Note that $|\vx|$ denotes the sum of elements of $\vx$ whereas $\norm{\vx}_1$
is the sum of absolute values of elements.

\begin{Lemma} The $m$-th power of a finite sum can be expanded as
  \cite{Apostol73},
  \begin{equation*}
    |\vx|^m = \sum_{|\vn|=m} \frac{m!}{\vn!} \vx^{\vn}.
  \end{equation*}
\end{Lemma}
\begin{proof}
  See \cite{Folland10}.
\end{proof}

\begin{Theorem}\label{th:1} The multivariate Taylor's expansion of a
  $(m+1)$-order differentiable function $f:\ \R^N\mapsto\R$ about the point
  $\vx\in\R^N$ is written as \cite{Apostol73},
  \begin{equation*}
    f(\vx+\vh) = f(\vx) + \sum_{|\vn|\leq m} \frac{\partial^{\vn}
      f(\vx)}{\vn!} \vh^{\vn} + \sum_{|\vn|=m+1} \frac{\partial^{\vn} f(\vx+
      t\,\vh)}{\vn!} \vh^{\vn}
  \end{equation*}
  for some $t\in(0,1)$.
\end{Theorem}
\begin{proof}
  See \cite{Folland10} and \cite{Apostol73}.
\end{proof}

\begin{Definition}
  A multivariate function $f(\vx)=f(x_1,\ldots,x_N)$ is said to be symmetric,
  if, for any permutation of its arguments $\vx$ denoted as $\vx^\prime$,
  $f(\vx)=f(\vx^\prime)$.
\end{Definition}

\section{Background Extensions}

In this section, additional results are obtained which are used in the next
section to introduce statistical sum-moments for random vectors. In particular,
the mean cosine similarity, the mean Minkowski distance as well as higher
central moments for random vectors are defined. A polynomial approximation of
univariate functions is shown to be a linear regression problem. A numerically
efficient solution of the LS problem is derived. Finally, a procedure to
generate multiple Gaussian processes with defined auto-covariance and
cross-covariance is devised.

Recall the second moment of a random variable $X$, i.e.,
\begin{equation*}
  \mu_2 = \E{ (X-c)^2 },\quad c\in\R.
\end{equation*} 
It is straightforward to show that $\mu_2$ is minimized for $c=\E{X}$, giving
the variance of $X$. On the other hand, $\mu_2=0$, if and only if
$c=\E{X}\pm\sqrt{-\var{X}}$.

For random vectors, both cosine similarity and Minkowski distance are random
variables. Assuming that the vectors are jointly stationary, and their elements
are identically distributed, the mean cosine similarity can be defined as,
\begin{equation*}
  \mScos(\vX_1,\vX_2) = \frac{\sum_{i=1}^N \E{X_{1i}X_{2i}}}
  {\sqrt{\sum_{i=1}^N \E{X^2_{1i}}-\vXb_{1i}^2}
    \sqrt{\sum_{i=1}^N X^2_{2i}-\vXb_{2i}^2}}
    = \frac{\sum_{i=1}^N \E{X_{1i}X_{2i}}}{N \sigma_1\sigma_2} =
    \frac{\Rb_{x_1x_2}}{\sigma_1\sigma_2}
\end{equation*}
where $\sigma_1^2=\var{X_1}$ and $\sigma_2^2=\var{X_2}$, and, the average
cross-correlation is calculated dimension-wise as,
\begin{equation*}
  \Rb_{x_1x_2} = \frac{1}{N} \sum_{i=1}^N \E{X_{1i}X_{2i}}.
\end{equation*}

The mean Minkowski distance for random vectors can be defined as,
\begin{equation}\label{eq:a8}
 \mSmnk(\vX_1,\vX_2) = \Big(\sum_{i=1}^N \E{\left| X_{1i}-X_{2i}\right|^m}
 \Big)^{1/m},\quad m=1,2,\ldots  
\end{equation}
Recognizing the $m$-th general moment in \eref{eq:a8}, the $m$-th power of the
mean Minkowski distance can be normalized as,
\begin{equation*}
  \tSmnk^m(\vX_1,\vX_2) = \sum_{i=1}^N \frac{\E{\left| X_{1i}-X_{2i}\right|^m}}
  {N \left(\E{\left| X_{1i}-X_{2i}\right|^2}\right)^{m/2}} =
  \bmu_m(\vX_1-\vX_2),\quad m=1,2,\ldots
\end{equation*}
where the average Minkowski distance between two random vectors is,
\begin{equation*}
  \bmu_m(\vX_1-\vX_2) = \frac{1}{N} \sum_{i=1}^N\ \mu_m(X_{1i}-X_{2i}),\quad
  m=1,2,\ldots
\end{equation*}

Furthermore, note that for $m=2$,
\begin{equation*}
  \frac{1}{2}\E{\norm{\vX_1-\vX_2}_2^2}+ \frac{1}{2}\E{\norm{\vX_1+\vX_2}_2^2}
  = \E{\norm{\vX_1}_2^2} + \E{\norm{\vX_2}_2^2}.  
\end{equation*}

Assuming positive integers $\vm=\{m_1,m_2,\ldots,m_N\}$, the higher order joint
central moments for a random vector $\vX=\{X_i\}_{1:N}$ can be defined as,
\begin{equation}\label{eq:z3}
  \mu_{m_1,\ldots,m_N}(X_1,\ldots,X_N) = \E{\prod_{i=1}^N (X_i-\Xb_i)^{m_i}}
\end{equation}
or, using more compact index notation as,
\begin{equation*}
  \mu_{\vm}(\vX) = \E{ (\vX-\vXb)^{\vm} }.
\end{equation*}

\subsection{Linear LS estimation}

The linear LS estimation can be used to fit a degree $(D-1)$ polynomial to $N$
samples of a random process $x(t)$ at discrete time instances
$t_1,t_2,\ldots,t_N$. Hence, consider the polynomial data model,
\begin{equation*}
  x(t) \approx  M(t;\vP) = \sum_{k=1}^D P_k t^{k-1}.
\end{equation*}
Denoting $\vw_i=[t_i^0,t_i^1,\ldots,t_i^{D-1}]^T$, the linear LS solution
\eref{eq:z0} gives the estimates,
\begin{equation*}
  \vPh = \left( \sum_{i=1}^N \left[
      \begin{array}{cccc} t_i^0 & t_i^1 & \cdots & t_i^{D-1} \\
        t_i^1 & t_i^2 & \cdots & t_i^{D} \\
        \vdots & \vdots & \ddots & \vdots \\
        t_i^{D-1} & t_i^D & \cdots & t_i^{2D-2} \\
      \end{array} \right] \right)^{-1} \sum_{i=1}^N x(t_i) \left[
    \begin{array}{c} t_i^0 \\ t_i^1 \\ \vdots \\ t_i^{D-1} \end{array} \right].
\end{equation*}

Assuming $D=2$ parameters, i.e., the linear LS regression for a straight line,
the parameters $P_1$ and $P_2$ to be estimated must satisfy the following
equality:
\begin{equation*}
  \frac{\df}{\df P_1} \sum_{i=1}^N \left( X_i - w_{1i} P_1 -w_{2i} P_2\right)^2
  = -2 \sum_{i=1}^N w_{1i} X_i + 2 \sum_{i=1}^N w^2_{1i} P_1 + 2 \sum_{i=1}^N
  w_{1i} w_{2i} P_2 \eqtext{!} 0.
\end{equation*}
Denoting the weighted averages, $\bX = \sum_{i=1}^N w_{1i} X_i $,
$\norm{\vw_1}_2^2 = \sum_{i=1}^N w^2_{1i}$, and
$\wb_{12} = \sum_{i=1}^N w_{1i} w_{2i}$, a necessary but not sufficient
condition for the linear LS estimation of parameters $P_1$ and $P_2$ is,
\begin{equation}\label{eq:z1}
  \bX = \norm{\vw_1}_2^2\, P_1 + \wb_{12} P_2.
\end{equation}
In the LS terminology, $(\norm{\vw_1}_2^2,\wb_{12})$ represent independent
variables whereas $\bX$ is a dependent variable.

Note that all $N$ measurements are used in \eref{eq:z1}. However, if $N$ is
sufficiently large, the data points could be split into two disjoint sets of
$N_1$ and $N_2$ elements, $N_1+N_2= N$, respectively. The parameters $P_1$ and
$P_2$ can be then readily estimated by solving the following set of two
equations, i.e.,
\begin{equation}\label{eq:w8}
  \left[ \begin{array}{c} \Xb_1=\frac{1}{a_1}\sum\limits_{i\in\II_1}\! w_{1i}
      X_i \\ \Xb_2=\frac{1}{a_2} \sum\limits_{i\in\II_2}\! w_{1i} X_i
      \\ \end{array} \right] = \left[
    \begin{array}{cc} \Wb_{11}= \frac{1}{a_1}\sum\limits_{i\in\II_1} \!
      w_{1i}^2 & \Wb_{12}= \frac{1}{a_1}\sum\limits_{i\in\II_1}\! w_{1i}w_{2i} \\
      \Wb_{21}=\frac{1}{a_2}\sum\limits_{i\in\II_2}\! w_{1i}^2 & \Wb_{22}=
      \frac{1}{a_2}\sum\limits_{i\in\II_2} \! w_{1i}w_{2i} \\\end{array} \right]
  \left[ \begin{array}{c} \Ph_1 \\ \\ \Ph_2 \\\end{array} \right]
\end{equation}
where $a_1,a_2>1$ are some constants (to be determined later), and $\II_1$ and
$\II_2$ are two disjoint index sets, such that
$\II_1\cup\II_2=\{1,2,\ldots,N\}$, and their cardinality, $|\II_1|=N_1$ and
$|\II_2|=N_2$. Note that,
\begin{equation}\label{eq:w3}
  \begin{array}{ccc}
    \norm{\vw_1}_2^2 &=& a_1\Wb_{11} + a_2\Wb_{21} \\
    \wb_{12} &=& a_1\Wb_{12} + a_2\Wb_{22} \\
    \bX &=& a_1\Xb_1 + a_2\Xb_2.
  \end{array}
\end{equation}
Consequently, the approximate LS estimates are computed as,
\begin{equation}\label{eq:w0}
  \left[ \begin{array}{c} \Ph_1 \\ \Ph_2 \\\end{array} \right] =
  \left[ \begin{array}{cc} \Wb_{11} & \Wb_{12} \\ \Wb_{21} & \Wb_{22} \\
    \end{array} \right]^{-1} \left[ \begin{array}{c} \Xb_1 \\ \Xb_2 \\
    \end{array} \right].
\end{equation}

\begin{figure}[H]
  \centering
  \includegraphics[scale=1.7]{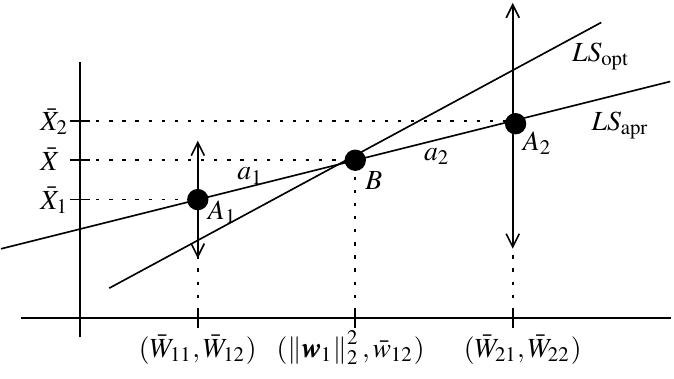}
  \caption{The exact ($\LSopt$) and the reduced complexity ($\LSapr$) linear LS
    regression.}\label{pict1}
\end{figure}

There are $2^N$ possibilities how to split $N$ data points into two disjoint
subsets indexed by $\II_1$ and $\II_2$. More importantly, the estimates
\eref{eq:w0} do not guarantee the minimum LS fit, i.e., achieving the minimum
squared error (MSE). However, the complexity of performing the LS fit is
greatly reduced by splitting the data, since independently of the value of
$N\gg 1$, only a $2\times 2$ matrix needs to be inverted. The optimum LS fit
and the approximate LS fit \eref{eq:w0} are depicted in \fref{pict1}. The
points $A_1$ and $A_2$ in \fref{pict1} correspond to the data indexes $\II_1$
and $\II_2$, respectively. The mid-point, $B=a_1A_1+a_2A_2$, follows from
\eref{eq:w3}. Note that $B$ is always located at the intersection of optimum
and approximate lines $\LSopt$ and $\LSapr$, respectively. The vertical arrows
at points $A_1$ and $A_2$ in \fref{pict1} indicate that the dependent values
$\Xb_1$ and $\Xb_2$ are random variables.

The larger the variation of the gradient of line $\LSapr$ in \fref{pict1}, the
larger the uncertainty and the probability that the line $\LSapr$ deviates from
the optimum line $\LSopt$. Given $0<p<1$, there exists $\xi>0$, such that the
probability of the gradient $\Gapr$ of $\LSapr$ to be within the given bounds
is,
\begin{equation}\label{eq:z2}
  \Prob{\bl(\xi) < \Gapr < \bu(\xi)}\geq p
\end{equation}
where
\begin{eqnarray*}
  \bl(\xi) &=& \frac{\left(\E{\Xb_2} - \xi\sqrt{\var{\Xb_2}}\right) -
    \left(\E{\Xb_1} + \xi\sqrt{\var{\Xb_1}}\right)}{\Wb_{22}-\Wb_{12}}\\
  \bu(\xi) &=& \frac{\left(\E{\Xb_2} + \xi\sqrt{\var{\Xb_2}}\right) -
    \left(\E{\Xb_1} - \xi\sqrt{\var{\Xb_1}}\right)}{\Wb_{22}-\Wb_{12}}.
\end{eqnarray*}
For stationary measurements (cf. \eref{eq:w8}), the means and the variances in
\eref{eq:z2} are equal to,
\begin{equation*}
  \begin{array}{ll}
    \E{\Xb_1} = \E{\Xb} N_1/(a_1N) & \var{\Xb_1} = \var{\Xb} N_1/(a_1^2 N) \\ 
    \E{\Xb_2} = \E{\Xb} N_2/(a_2N) & \var{\Xb_2} = \var{\Xb} N_2/(a_2^2N).
  \end{array}
\end{equation*}

The uncertainty in computing $\Gapr$ from data, and thus, also the probability
of line $\LSapr$ deviating from line $\LSopt$ is inversely proportional to the
width of the interval in \eref{eq:z2}, i.e.,
\begin{equation}\label{eq:v8}
  \bu(\xi)-\bl(\xi) = \frac{2\xi( \sqrt{\var{\Xb_1}}+ \sqrt{\var{\Xb_2}} )}
  {\Wb_{22}-\Wb_{12}} \propto \frac{\frac{\sqrt{N_1}}{a_1}+
    \frac{\sqrt{N_2}}{a_2}}{\Wb_{22}-\Wb_{12}}.
\end{equation}
Consequently, the numerator of \eref{eq:v8} must be minimized, and the
denominator maximized.

In order to minimize the numerator in \eref{eq:v8}, it is convenient to choose,
$a_1=N_1^\nu$ and $a_2=N_2^\nu$, where $\nu\in\Rp$ is a constant to be
optimized. It is straightforward to show that the expression,
$(N_1^{1/2-\nu}+N_2^{1/2-\nu})$ is convex, i.e., it has a unique global minimum
for $\forall \nu>1/2$. Hence, a necessary condition to reduce the approximation
error is that $a_1>N_1^{1/2}$ and $a_2>N_2^{1/2}$. For numerical convenience,
let $a_1=N_1$ and $a_2=N_2$. Then, the definitions of dependent and independent
variables in \eref{eq:w8} become arithmetic averages, and the optimum index
subsets have the cardinality,
\begin{equation}\label{eq:e3}
  \left\{ \begin{array}{cc} |\II_1|=|\II_2|=N/2 & N-\mbox{even} \\
      |\II_1|=|\II_2|\pm 1= (N\pm1)/2 & N-\mbox{odd}.
    \end{array} \right.
\end{equation}

Furthermore, in order to maximize the denominator in \eref{eq:v8}, assume that
the independent variables $(\norm{\vw_1}_2^2,\wb_{12})$ in \eref{eq:w8} are
sorted by $w_{li}$, i.e., let $w_{11}< w_{12}< \ldots < w_{1N}$. This ordering
and the condition \eref{eq:e3} suggests that the disjoint index sets $\II_1$
and $\II_2$ maximizing the difference, $\Wb_{22}-\Wb_{12}$, are,
\begin{equation*}
  \left\{ \begin{array}{cc} \II_1=\{1,2,\ldots,N/2\}, \II_2=\{N/2+1,\ldots,N\}
      & N-\mbox{even} \\ 
      \II_1=\{1,2,\ldots,(N-1)/2\},\II_2=\{(N+1)/2,\ldots,N\} & N-\mbox{odd}\\
      \mbox{or, } \II_1=\{1,2,\ldots,(N+1)/2\},\II_2=\{(N+3)/2,\ldots,N\}.
    \end{array} \right.
\end{equation*}

In summary, the approximate linear LS regression can be efficiently computed
with a good accuracy by splitting the data into multiple disjoint subsets,
calculating the average data points in each of these subsets, and then solving
the set of linear equations with the same number of unknown parameters. The
data splitting should exploit ordering of data points by one of the independent
variables. Moreover, it can be expected that the accuracy of the approximate LS
regression is going to improve with the number of data points $N$. Numerical
evaluation of the approximate LS regression is considered in Section 5.

\subsection{Generating pairwise-correlated Gaussian processes}

How to generate a single correlated Gaussian process is well established in the
literature, and it has been described in Section 2.3. Note also that the linear
transformation matrix assumed in Section 2.3 does not have to be square.
\begin{Lemma}
  The matrix, $\vT^T\vT$, $\vT\in\R^{N_1\times N_2}$, is positive definite,
  provided that, $N_1\leq N_2$.
\end{Lemma}
\begin{proof}
  The matrix is positive definite, if $\vU^T\vT^T\vT\vU>0$ for
  $\forall \vU\in\R^{N_2}$. The product,
  $\vU^T\vT^T\vT\vU = (\vT\vU)^T(\vT\vU) = \norm{\vT\vU}_2^2>0$ where
  $\norm{\cdot}_2$ is the Euclidean norm of a vector. If $N_1>N_2$, then some
  rows in $\vT$ must be linearly dependent, and it is then possible to find a
  non-zero vector $\vU$, such that $\vT\vU = 0$.
\end{proof}
Consequently, a linear transformation of uncorrelated Gaussian vector
$\vU\in\R^{N_2}$ can be used to generate a correlated Gaussian vector
$\vX\in\R^{N_1}$, provided that, $N_1\leq N_2$.

Furthermore, it is often necessary to generate multiple mutually correlated
Gaussian processes with given auto-correlation and cross-correlation.
\begin{Lemma}
  The linear transformation,
  \begin{equation}\label{eq:v9}
    \left[ \begin{array}{c} \vx_1 \\ \vx_2 \end{array} \right] =
    \left[ \begin{array}{cc} \vT_1 & \vK \\ \vZs & \vT_2 \end{array} \right]
    \left[ \begin{array}{c} \vu_1 \\ \vu_2  \end{array} \right] 
  \end{equation}
  generates a pair of correlated Gaussian vectors $\vx_1\in\R^{N_1}$ and
  $\vx_2\in\R^{N_2}$ from uncorrelated zero-mean Gaussian vectors
  $\vu_1,\vu_2\in\R^{N}$ where $\vZs$ denotes a zero matrix, and according to
  Lemma 8, it is necessary that, $\max(N_1,N_2)\leq N$. The corresponding
  (auto-) correlation and cross-correlation matrices are,
  \begin{equation*}
    \E{\vu_1\vu_1^T} = \sigma_1^2 \vI,\quad \E{\vu_2\vu_2^T} =
    \sigma_2^2 \vI, \quad \E{\vu_1\vu_2^T} = \vZs
  \end{equation*}
  \begin{equation*}
    \vC_{x_1}=\E{\vx_1\vx_1^T} = \vT_1\vT_1^T+\vK\vK^T,\quad
    \vC_{x_2}=\E{\vx_2\vx_2^T} = \vT_2\vT_2^T,\quad \vC_{x_1x_2} =
    \E{\vx_1\vx_2^T} = \vK\vT_2^T = \vC_{x_2x_1}^T
  \end{equation*}
  where $\vI$ denotes an identity matrix.
\end{Lemma}
\begin{proof}
  The proof is straightforward by substituting \eref{eq:v9} to definitions of
  $\vC_{x_1}$, $\vC_{x_2}$ and $\vC_{x_1x_2}$.
\end{proof}

The following corollary details the procedure described in Lemma 9.
\begin{Corollary}
  Given $\vT_2$, calculate the (auto-) correlation matrix,
  $\vC_{x_2}=\vK\vT_2^T$, or vice versa. Then, given the cross-correlation
  matrix, $\vC_{x_1x_2}$, compute $\vK= \vC_{x_1x_2} \vT_2 \vC_{x_2}^{-T}$.
  Finally, given $\vT_1$, calculate the (auto-) correlation matrix,
  $\vC_{x_1}= \vT_1\vT_1^T+\vK\vK^T$, or, obtain $\vT_1$ by solving the matrix
  equation, $\vT_1\vT_1^T = \vC_{x_1} - \vK\vK^T$. Note that the matrix
  equation, $\vT^T\vT=\vC$, can be solved for $\vT$ using the singular value
  decomposition, $\vC = \vU\,\vLam\,\vU^T$ where $\vU$ is a unitary matrix and
  $\vLam$ is a diagonal matrix of eigenvalues. Then, $\vT= \vU\sqrt{\vLam}$.
\end{Corollary}

\section{Polynomial Statistics and Sum-Moments for Vectors of Random Variables}

The main objective of this section is to define a universal function to
effectively measure the statistics of random vectors and random processes
observed in multiple discrete time instances. The function should: (1) be
universally applicable for an arbitrary number of random variables and random
vectors, (2) be symmetric, so that all random variables are considered equally,
(3) lead to mathematically tractable expressions, and (4) be convex, so its
parameters can be optimized.

Let $f:\ \R^N\mapsto\R$ denote such a mapping from $N$ random variables $X_i$
to a scalar random variable $Y$, i.e.,
\begin{equation}\label{eq:x1}
  Y = f ( X_1,X_2,\ldots,X_N ) = f( \vx )
\end{equation}
where, for a random process, $X_i=x(t_i)$. In order to satisfy the symmetry
requirement, it can be assumed that the random variables $X_i$ are first
combined as,
\begin{equation}\label{eq:q7}
  Y = f ( X_1 \circ X_2 \circ \cdots \circ X_N ) 
\end{equation}
using a binary commutative operator $\circ$ such as addition or multiplication.
More importantly, if the random variables in \eref{eq:q7} are combined using
addition, then the function $f$ must be non-linear.

The mapping \eref{eq:x1} or \eref{eq:q7} defines a scalar random field over
discrete time instances corresponding to the discrete time samples of the
random process $x(t)$.
\begin{Definition}
  A random process $x(t)$ observed at $N$ discrete time instances
  $\vt=\{t_i\}_{1:N}$ defines a $N$-dimensional scalar random field,
  \begin{equation*}
    Y = \cF( t_1,t_2,\ldots,t_N ) = \cF(\vt) = f( x(\vt) )  
  \end{equation*}
  where the index notation, $x(\vt)=\{X_1=x(t_1),\ldots,X_N=x(t_N)\}$.
\end{Definition}
Assuming a multivariate Taylor's expansion of the function $f(\vx)$ defined in
Theorem 1 on \pref{th:1} about its mean $\vxb=\E{\vx}$, the random field can be
expressed as,
\begin{equation}\label{eq:x6}
  f(\vxb+\vh) = f(\vxb) + \sum_{l=1}^m \ \sum_{|\vn|=l} \frac{\partial^{\vn}
    f(\vx)}{\vn!} \vh^{\vn} + R_m
\end{equation}
where $R_m$ is the reminder. Thus, given $\vxb$, the value of $f(\vxb+\vh)$ is
a weighted sum of $\vh^{\vn}$ plus an offset $f(\vxb)$.

The key realization is that any function $f$ having all the desired properties
can be defined by combining the equations \eref{eq:q7} and \eref{eq:x6}. In
particular, neglecting the reminder $R_m$, the number of parameters in
\eref{eq:x6} can be greatly reduced by approximating the weights
$\partial^{\vn} f(\vx)$ in \eref{eq:x6} with the coefficients $(l!\,p_l)$ which
are independent of $\vn$. Moreover, instead of precisely determining the values
of $p_l\in\R$ to obtain the best possible approximation of the original
function $f$, it is useful as well as sufficient to constrain the Taylor
expansion \eref{eq:x6} for the class of functions that are exactly constructed
as,
\begin{equation*}
  f(\vxb+\vh) = f(\vxb) + \sum_{l=1}^m  p_l \sum_{|\vn|=l} \frac{l!}{\vn!}
  \vh^{\vn} = \sum_{l=0}^m p_l (h_1+\cdots+h_N)^l
\end{equation*}
using Lemma 6, and denoting $p_0= f(\vxb)$.
\begin{Lemma}
  A function $f:\ \R^N\mapsto\R$ having the desirable properties to measure
  statistics of a random vector $\vx$ is the $m$-th degree polynomial,
  \begin{equation}\label{eq:x8}
    Y = f(\vx) = \tilde{f}(|\vx|) = \sum_{l=0}^m p_l |\vx-\vxb|^l =
    \sum_{l=0}^m p_l \left( \sum_{i=1}^N (X_i-\Xb_i) \right)^l
  \end{equation}
  where $p_0= \vxb = \E{\vx}$, and the value $m$ and the coefficients
  $p_l\in\R$ are determined by the requirements of a given application.
\end{Lemma}
\begin{proof}
  The proof would repeat the results and discussion given in the text
  preceding the lemma. 
\end{proof}

The most important statistical property of the random variable $Y=f(\vx)$ is
its mean value.
\begin{Lemma}
  The mean value of the random field $\cF(\vt)$ created from $N\leq K$ discrete
  time observations of the $K$-th order stationary random process $x(t)$ is a
  $(N-1)$-dimensional scalar field,
  \begin{equation*}
    \Yb = \E{\cF(\vt)} = \cC(t_2-t_1,t_3-t_1,\ldots,t_N-t_1) =
    \cC(\tau_1,\tau_2,\ldots,\tau_{N-1}) = \cC( \vtau )
  \end{equation*}
  where $\tau_i=t_{i+1}-t_1$, $i=1,2,\ldots,N-1$.
\end{Lemma}
\begin{proof}
  Assuming the polynomial field constructor $f$ defined in Lemma 10, its mean
  value is equal to the weighted sum of central moments,
  $\mu_l(\vx)=\E{ |\vx-\vxb|^l }$. According to Lemma 1, the condition
  $N\leq K$ is sufficient to reduce the dimension of expectation by one. For
  example, choosing the first sample $X_1=x(t_1)$ as a reference, the
  probability density of samples becomes,
  $f_X(\vx;0, t_2-t_1,\ldots,t_N-t_1) \equiv
  \tilde{f}_X(\vx;t_2-t_1,\ldots,t_N-t_1)$.
\end{proof}

\subsection{Related concepts}

Assume the scalar field constructor $f$ defined in \eref{eq:x8} for a $K$-th
order stationarity random process $x(t)$. Define the auxiliary random variable,
\begin{equation*}
  Z(a) = \frac{1}{a} \sum_{i=1}^N (X_i-\Xb_i)
\end{equation*}
where $a$ is a normalization constant, $a\neq 0$. The expression \eref{eq:x8}
can be then rewritten as, $Y(a)=\sum_{l=0}^m p_l\, Z^l(a)$. For $a=1$, $Z(1)$
has a zero mean, and the variance, $\var{Z(1)}\leq \sum_{i=1}^N \var{X_i}$,
since $X_i$ are, in general, correlated. For $a=N$, $Z(N)$ represents a sample
mean. In this case, the variance of $Z(N)$ is reduced to,
$\var{Z(N)}\leq \var{X_i}/N$. For $a=\sqrt{N}$, the variance of $Z(\sqrt{N})$
becomes independent of $N$ and $\var{Z(\sqrt{N})}\leq \var{X_1}$.

For $p_l=s^l/l!$, in the limit of large $m$, eq. \eref{eq:x8} gives,
\begin{equation*}
  Y = \lim_{m\to\infty}\ \sum_{l=0}^m\ \frac{s^l}{l!}\, (Z(a))^l =
  \eee^{s\,Z(a)}.
\end{equation*}
The mean, $\E{Y}=\E{\eee^{s\,Z(a)}}$, is the moment generating function of
random variable $Z(a)$.

In data processing, the sample mean is intended to be an estimate of the true
population mean, i.e., $N\gg 1$ is required. Here, $Z(a)$ is calculated over a
finite number of vector or process dimensions, so it is a random variable for
$\forall a\in\R\setminus\{0\}$. The variable $Z(N)$ should be then referred to
as arithmetic average or a center of gravity of the random vector $\vX$ in the
Euclidean space $\R^N$, i.e.,
\begin{equation}\label{eq:x5}
  Z(N)\triangleq\Xb = \frac{1}{N}\sum_{i=1}^N X_i\in\R.
\end{equation}
Note that \eref{eq:x5} is not an $l_1$-norm, since the variables $X_i$ are not
summed with absolute values.

If random variables $X_i$ are independent, the distribution of $Z(a)$ is given
by convolution of their marginal distributions. For correlated observations, if
the multivariate characteristic function,
$\ft(\vs) = \E{\eee^{\jj \vs\cdot \vX}}$, of $\vX$ can be obtained, the
distribution of $Z(a)=|\vX|/a$ is calculated as,
\begin{equation*}
  f_Z(Z(a)) = \frac{1}{2\pi} \int_{\R} \eee^{-\jj s Z(a)}
  \ft\left(\frac{s}{a}\cdot\vOs\right)\df s
\end{equation*}
where the dot product, $\frac{s}{a}\cdot\vOs=<\frac{s}{a},\vOs>$, and $\vOs$
denotes an all-ones vector.

Many other properties involving the sums of random variables can be obtained.
For instance, if the random variables $X_i$ are independent and have zero mean,
then,
\begin{equation*}
  \E{ Z^m(1) } - \E{ (Z(1) - X_N)^m } =  \left\{ \begin{array}{cc}
      \E{X_N^m} & m=2,3 \\ \E{X_N^4} + 6 \sum_{i=1}^{N-1} \E{X_N^2}\E{ X_i^2} &
      m=4 \\ \E{X_N^5} + 10 \sum_{i=1}^{N-1} \E{X_N^2}\E{X_i^3} +
      \E{X_N^3}\E{X_i^2} & m=5. \\\end{array} \right.
\end{equation*}

Considering Lemma 10 and 11, an important statistic for a random vector can be
defined as the $m$-th central sum-moment.
\begin{Definition}
  The $m$-th central sum-moment of a random vector $\vX\in\R^N$ is computed as,
  \begin{equation*}
    \sum\mu_m(\vX) = \mu_m(|\vX|) =  \E{\, \left| \sum_{i=1}^N
        (X_i-\Xb_i) \right|^m\, },\quad m=1,2,\ldots
  \end{equation*}
\end{Definition}
\begin{Lemma}
  The second central sum-moment of a random vector is equal to the sum of all
  elements of its covariance matrix, i.e.,
  \begin{equation*}
     \sum\mu_2(\vX) = \mu_2(|\vX-\vXb|) = \E{ \left( \sum_{i=1}^N
        (X_i-\Xb_i) \right)^2\, } = \sum_{i,j=1}^N \cov{X_i,X_j} = |\vC_x|.
  \end{equation*}
  Furthermore, the second central sum-moment is also equal to the variance of
  $|\vX|$, i.e.,
  \begin{equation*}
    \sum\mu_2(\vX) = \var{\,|\vX|\,} = \var{\sum_{i=1}^N (X_i-\Xb_i)}.
  \end{equation*}
\end{Lemma}
\begin{proof}
  The first equality is shown by expanding the expectation, and substituting
  for elements of the covariance matrix, $[\vC_x]_{i,j}= \cov{X_i,X_j}$. The
  second expression follows by noting that $\sum_{i=1}^N (X_i-\Xb_i)$ has zero
  mean.
\end{proof}

In the literature, there are other measures involving sums of random variables.
For instance, in Mean-Field Theory, the model dimensionality is reduced by
representing $N$-dimensional vectors by their center of gravity
\cite{Yedidia01}. The central point of a vector is also used in the first order
approximation of multivariate functions in \cite{Sobol01}, and in the model
overall variance in \cite{Saltelli05}.

In Measure Theory \cite{Shirali18}, the total variation (TV) of a real-valued
univariate function, $x: (t_0,t_N)\mapsto\R$, is defined as the supremum over
all $N$-segment partitioning of the interval $(t_0,t_N)$, i.e.,
\begin{equation*}
  \mathrm{TV}(x) = \sup\limits_{t_i\in(t_0,t_N)} \ \sum_{i=0}^{N-1}
  |x(t_{i+1}-x(t_i)|.
\end{equation*}
The TV concept can be adopted for observations $X_i=x(t_i)$ of a stationary
random process $x(t)$ at discrete time instances, $\{t_i\}_{0:N}$. A
mathematically tractable mean TV measure can be defined as,
\begin{equation*}
  \overline{\mathrm{TV}}^2(x) = \E{ \sum_{i=0}^{N-1} |X_{i+1}-X_i|^2 } =
  2 N (\E{X_i^2} - \cov{X_{i+1},X_i}).
\end{equation*}

Jensen's inequality for a random vector assuming equal weights can be stated
as \cite{Boyd04},
\begin{equation*}
  \E{ \left|\frac{1}{N} \sum_{i=1}^N (X_i-\Xb_i) \right|^m } \leq
  \frac{1}{N} \sum_{i=1}^N \E{|X_i-\Xb_i|^m}.
\end{equation*}
Alternatively, exchanging the expectation and summation, Jensen's inequality
becomes,
\begin{equation}\label{eq:x7}
  \sum_{i=1}^N \big|\E{ X_i-\Xb_i}\big|^m  \leq \E{ \sum_{i=1}^N
    |X_i-\Xb_i|^m}.
\end{equation}
Moreover, if the right-hand side of \eref{eq:x7} is to be minimized and $m=2$,
the inequality in \eref{eq:x7} changes to equality. In particular, consider the
minimum mean square error (MMSE) estimation of a vector of random parameters
$\vP=\{P_i\}_{1:N}$ from measurements $\vX$. Denoting $\Pb_i(\vX)=\E{P_i|\vX}$,
the MMSE estimator $\vPh(\vX)$ minimizes \cite{Kay93},
\begin{eqnarray}
  \min_{\Ph}\ \E{\sum_{i=1}^N (\Ph_i(\vX)-P_i)^2} &=& \min_{\Ph}\ \E{
    \sum_{i=1}^N \left((\Ph_i(\vX)-\Pb_i(\vX)) - (P_i-\Pb_i(\vX))\right)^2 }
  \nonumber \\ &=& \min_{\Ph} \left\{ \sum_{i=1}^N (\Ph_i(\vX)-\Pb_i(\vX))^2 +
    \E{\sum_{i=1}^n (P_i-\Pb_i(\vX))^2} \right\} \label{eq:y4} \\
  &=& \min_{\Ph}\ \sum_{i=1}^N (\E{\Ph_i(\vX)-P_i|\vX})^2 =
  \min_{\Ph}\ \sum_{i=1}^N (\Ph_i(\vX)-\E{P_i|\vX})^2  \nonumber
\end{eqnarray}
where the expectations are over the conditional distribution $f_{\vP|\vX}$.

In signal processing, a length $N$ moving average (MA) filter transforms the
input sequence $X_i$ into an output sequence $Y_i$ by discrete-time convolution
$\circledast$, i.e.,
\begin{equation*}
  Y_i = \sum_{j=0}^{N-1} X_{i-j} = [\underbrace{1\ 1\ \ldots\ 1}_{\vOs_N}]
  \circledast X_i.
\end{equation*}
The (auto-) correlations of the input and output sequences are related by Lemma
3 on \pref{lm:3}, i.e.,
\begin{equation}\label{eq:u3}
  C_Y(i) = \vOs_N \circledast \vOs_N \circledast C_x(i) =
  \sum_{j=-N+1}^{N-1} (N-|j|)\, C_x(i-j).
\end{equation}
Note that if the input process is stationary, then the input and output
processes are jointly stationary.

\subsection{Multiple random processes}

The major complication with observing, evaluating and processing multiple
random processes is how to achieve their time alignment and amplitude
normalization (scaling). Focusing here on time alignment problem only, denote
the discrete time observation instances of $L$ random processes as,
\begin{equation*}
  \vt_l = \{ t_{l1} < t_{l2} < \ldots < t_{lN_l} \},\quad l=1,2,\ldots,L
\end{equation*}
Assume that the first time instance $t_{l1}$ of every process serves as a
reference. Then, there are $(L-1)$ uncertainties in time alignment of $L$
processes, i.e.,
\begin{equation*}
  \Delta_l = (t_{l1} - t_{11})\in\R, \quad l=2,3,\ldots,L
\end{equation*}
The $(L-1)$ values $\Delta_l$ are unknown parameters which must be estimated.
Note also that the difference,
\begin{equation*}
  \Delta_l - \Delta_k = t_{l1} - t_{k1}
\end{equation*}
represents an unknown time shift between the process $x_l(t)$ and $x_k(t)$.

For any multivariate stationary distribution of observations of random
processes, the corresponding cross-correlation normally attains a maximum for
some time shift between any two processes \cite{Gardner90}. Hence, a usual
strategy for aligning the observed sequences is to locate the maximum value of
their cross-correlation. The time shifts $\Delta_l$, $l=1,2,\ldots,L$, are then
estimated as,
\begin{equation*}
  \hat{\Delta}_l = \argmax_\Delta C_{x_1x_l}(\Delta),\quad
  \Delta\in\{(t_{li} - t_{11})\}_{i=1,\ldots,N_l}.  
\end{equation*}

Assuming the center values $\Xb_1$ and $\Xb_2$ defined in \eref{eq:x5} as
scalar representations of vectors $\vX_1$ and $\vX_2$, their cross-covariance
can be computed as,
\begin{equation}\label{eq:y1}
  N^2 \cov{\Xb_1,\Xb_2} = \cov{|\vX_1|,|\vX_2|} = N^2\, \E{
    (\Xb_1-\E{X_1})(\Xb_2-\E{X_2}) }.
\end{equation}
The task is now how to generalize the pairwise cross-covariance \eref{eq:y1}
for the case of multiple random vectors having possibly different lengths. If
all random vectors of interest are concatenated into one single vector, the
$m$-th joint central sum-moment can be then defined by utilizing Lemma 10 and
Lemma 11.
\begin{Definition}
  The $m$-th central sum-moment for $L$ random processes with $N_l$
  observations, $l=1,2,\ldots,L$, is computed as,
  \begin{equation*}
    \sum\mu_m(\vX_1,\ldots,\vX_L) = \mu_m(|\vX_1|+\ldots+|\vX_L|) =
    \E{\, \left| \sum_{l=1}^L \sum_{i=1}^{N_l} (X_{li}-\Xb_{li}) \right|^m\, }.
  \end{equation*}
\end{Definition}
\begin{Lemma}
  The second central sum-moment for $L$ random processes with $N_l$
  observations, $l=1,2,\ldots,L$, is equal to the sum of all pairwise
  covariances, i.e.,
  \begin{equation*}
    \sum\mu_2(\vX_1,\ldots,\vX_L) = \sum_{l,k=1}^L \sum_{i,j=1}^{N_l} \cov{
      X_{li},X_{kj}} = \sum_{l,k=1}^L |\cov{\vX_l,\vX_k}| =
    \var{|\vX_1|+\ldots+|\vX_L|}.
  \end{equation*}
\end{Lemma}
\begin{proof}
  The expression can be obtained by expanding the sum, and then applying the
  expectation.
\end{proof}

Many other properties of central and non-central sum-moments can be obtained.
For example, assuming two equal-length vectors $\vX_1$ and $\vX_2$, it is
straightforward to show that,
\begin{eqnarray*}
  \E{\left(|\vX_1|+|\vX_2|\right)^2} - \E{|\vX_1|^2 + |\vX_2|^2} &=&
  2\sum_{i,j=1}^N \E{X_{1i}X_{2j}} \\
  \E{\left(|\vX_1|+|\vX_2|\right)^2} -  \E{\norm{\vX_1-\vX_2}_2^2} &=&
  \sum_{\substack{i,j=1 \\ i\neq j}}^N \E{ X_{1i}X_{1j} +
    X_{2i}X_{2j}} + 2 \sum_{i,j=1}^N \E{ X_{1i}X_{2j}} + 2 \sum_{i=1}^N
  \E{X_{1i} X_{2i}}.   
\end{eqnarray*}

\section{Illustrative Examples}

This section provides examples to quantitatively evaluate the results developed
in the previous sections. In particular, the accuracy of approximate linear LS
regression proposed in Section 3.1 is assessed to justify its lower
computational complexity. The central sum-moments introduced in Section 4 are
compared assuming correlated Gaussian processes. Finally, several signal
processing problems involving the 1st order Markov process are investigated.

\subsection{Linear regression}

Consider a classical one-dimensional linear LS regression problem with
independent and identically normally distributed errors. The data points are
generated as,
\begin{equation*}
  X_i = \Delta i:\ Y_i = P_2 X_i + P_1+E_i,\quad i=1,2,\ldots,N
\end{equation*}
where $E_i$ are zero-mean, uncorrelated Gaussian samples having the equal
variance $\sigma_e^2$, and $P_1$ and $P_2$ are unknown parameters to be
estimated. This LS problem can be solved exactly using the expression
\eref{eq:z0}, and substituting $w_{1i}=1$ and $w_{2i}=\Delta i$,
$\forall i=1,2,\ldots,N$. Alternatively, to avoid inverting the $N\times N$
data matrix, the procedure devised in Section 3.1 suggests to split the data
into two equal-size subsets, compute the average data point for each subset,
and then solve the corresponding set of two equations with two unknowns.
Specifically, the set of two equations with unknown parameters $\Ph_1$ and
$\Ph_2$ is,
\begin{equation}\label{eq:u7} 
  \begin{array}{ccc}
    \frac{2}{N} \sum_{i=1}^{N/2} Y_i &=& \Ph_1 + \Ph_2\, \frac{\Delta}{8}N(2+N)
    \\ \frac{2}{N} \sum_{i=N/2+1}^N Y_i &=& \Ph_1 + \Ph_2\,
    \frac{\Delta}{8}N(2+3N)
  \end{array}
\end{equation}
assuming $N$ is even, and using,
$\sum_{i=1}^{N/2}\Delta i = \frac{\Delta}{8}N(2+N)$, and
$\sum_{i=N/2+1}^{N}\Delta i= \frac{\Delta}{8}N(2+3N)$. Denoting
$\bY_1= \frac{2}{N}\sum_{i=1}^{N/2} Y_i$ and
$\bY_2 = \frac{2}{N} \sum_{i=N/2+1}^N Y_i$, the closed-form solution of
\eref{eq:u7} is,
\begin{equation*}
  \begin{array}{ccl}
    \Ph_1 &=& \frac{1}{2N}\left((2+3N)\bY_1 - (2+N) \bY_2\right)\\
    \Ph_2 &=& \frac{4}{\Delta N^2} (\bY_2 - \bY_1).
  \end{array} 
\end{equation*}

\begin{figure}[H]
  \centering
  \psfrag{xx}[][][1.0]{$N_1$}
  \psfrag{yy}[][][1.0]{$\bT\pm\sqrt{\var{T}}$}
  \includegraphics[scale=0.6]{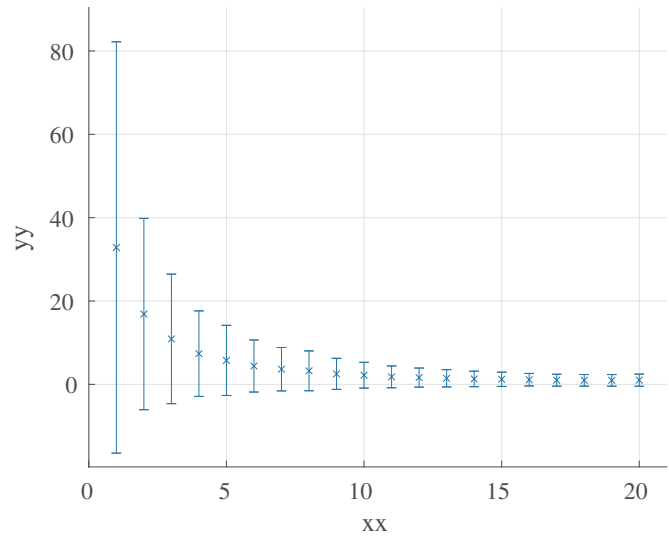}
  \caption{The relative total mean-square error of the approximate linear LS
    regression.}\label{pict2}
\end{figure}

As a numerical example, assume the true values $P_1=1.5$, $P_2=0.3$,
$\E{E_i}=0$, $\var{E_i}=1$, and $N=40$ data points. \fref{pict2} shows the
intervals $(\bT-\sqrt{\var{T}}, \bT+\sqrt{\var{T}})$ versus the subset size
$1\leq N_1\leq N/2$ for the random variable $T$ defined as,
\begin{equation*}
  T = 100\ \frac{\Sapr(N)-\Sopt(N)}{\Sopt(N)}
\end{equation*}
where $\Sapr(N)=\sum_{i=1}^N (Y_i-\Ph_1-\Ph_2 X_i)^2$ and
$\Sopt(N)=\sum_{i=1}^N (Y_i-P_1-P_2 X_i)^2$ are the total MSEs. In the limit,
$\lim_{N\to\infty} (\Sapr(N) - \Sopt(N)) = 0$, since a sufficiently large
subset of data is as good as the complete set of data. For finite $N$, it is
likely that $\Sapr(N)>\Sopt(N)$, so the lower bounds in \fref{pict2} converge
much faster to zero than the upper bounds.

\subsection{Comparison of central moments}

\begin{figure}[H]
  \centering
  \psfrag{xx}[][][0.8]{$N$}
  \psfrag{yy}[][][0.8]{$\sum \mu_2(\vX)$}
  \includegraphics[scale=0.55]{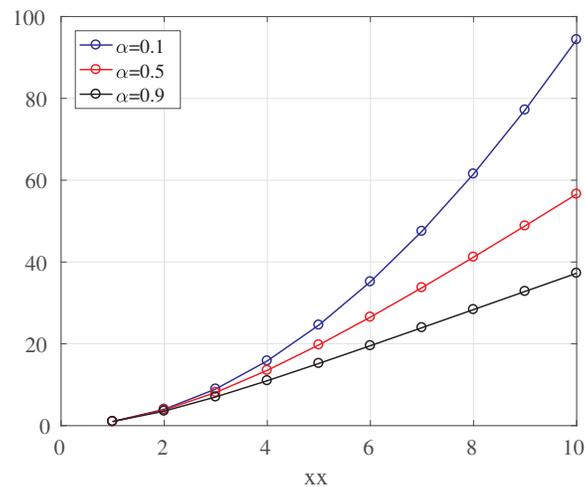}
  \caption{The second central sum-moments of the 2nd order Markov process with
    parameter $\alpha$ and length $N$.}\label{pict4}
\end{figure}

Assuming Lemma 12 and eq. \eref{eq:x9}, the second central sum-moment of the
2nd order Markov process of length $N$ is,
\begin{equation*}
  \sum \mu_2(\vX) = \sum_{i,j=1}^N C_{\MPP}(i-j) = \sigma^2 \left(N +
    2 \sum_{i=1}^{N-1} i (1+(N-i)\alpha) \eee^{-\alpha(N-i)}\right).
\end{equation*}
These moments are compared in \fref{pict4} for different values of the sequence
length $N$, three values of the parameter $\alpha$, and $\sigma^2=1$. It can be
observed that the values of the second central sum-moment are increasing with
$N$ and the level of correlation, $\eee^{-\alpha}$.

Consider now the following three central moments of order $m=1,2,\ldots$, i.e.,
\begin{eqnarray*}
  \mSmnk(N) &=& \sum_{i=1}^N \E{|\sqrt{N}\, X_i|^m} \\
  \sum\mu_2(N) &=&  \E{\left|\sum_{i=1}^N X_i \right|^m} \\
  \sum\tilde{\mu}_2(N) &=& \E{\left( \sum_{i=1}^N |X_i| \right)^m}.
\end{eqnarray*}
The moment, $\mSmnk$, is the mean Minkowski distance; the scaling by $\sqrt{N}$
is introduced to facilitate the comparison with the other two moments, i.e.,
the mean sum-moment $\sum\mu_2$, and the mean sum-moment $\sum\tilde{\mu}_2$
having the samples summed as the $l_1$ norm. More importantly, assuming a
correlated Gaussian process $X_i=x(t_i)$, the central sum-moment can be readily
obtained in a closed form whereas obtaining the closed form expressions for the
other two metrics may be mathematically intractable. In particular, by
factoring the covariance matrix as, $vC_x=\vT_x\vT_x^T$, the correlated
Gaussian vector can be expressed as, $\vX=\vT\vU$. Then the sum,
$|\vX|= \vOs^T\vT\vU$, and the $m$-th central sum-moment can be computed as,
\begin{equation*}
  \sum\mu_2(N) = \E{ \left| \vOs^T\vT\vU \right|^m } =
  \norm{\vOs^T\vT}_2^m\, \E{ | U |^m } = \norm{\vOs^T\vT}_2^m\,
  \frac{2^{m/2}}{\sqrt{\pi}} \Gamma\!\left( \frac{m+1}{2}\right)
\end{equation*}
where $U$ is a zero-mean, unit-variance Gaussian random variable, and $\Gamma$
denotes the gamma function.

\begin{figure}[H]
  \centering
  \psfrag{xx}[][][0.8]{$N$}
  \psfrag{yy}[][][0.8]{$\mSmnk/N$, $\sum\mu_2/$, $\sum\tilde{\mu}_2/N$}
  \psfrag{zz1}[][][0.8]{$\alpha=0.2$}
  \psfrag{zz2}[][][0.8]{$\alpha=0.5$}
  \psfrag{zz3}[][][0.8]{$\alpha=0.8$}
  \includegraphics[scale=0.37]{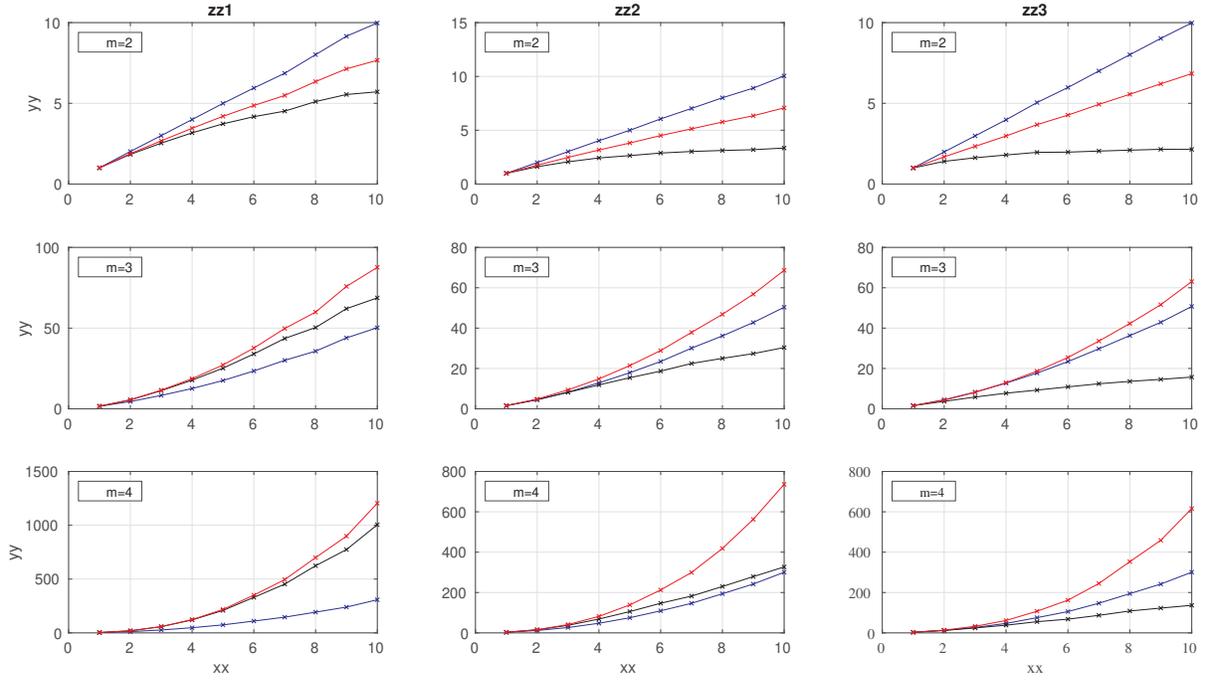}
  \caption{The Minkowski (blue), sum-moment (black) and sum-moment of absolute
    values (red) mean statistics for the 1st order Markov sequence of length
    $N$. Columns: different values $\alpha$. Rows: different values
    $m$.}\label{pict5}
\end{figure}

\fref{pict5} shows all three moments as a function of sequence length $N$ for
three values of parameter $\alpha$ assuming the 1st order Markov process. The
vertical axis in \fref{pict5} is scaled by $1/N$ for convenience. Note that,
for uncorrelated, i.e., independent Gaussian samples, the moments $\sum\mu_2$
and $\sum\tilde{\mu}_2$ are identical. More importantly, all three moments are
strictly increasing with the number of samples $N$ and with the order $m$.

\subsection{Signal processing problems for the 1st order Markov process}

Consider the 1st order Markov process observed at the output of a length $N$ MA
filter. According to \eref{eq:u3}, the (auto-) covariance of the output process
is,
\begin{equation}\label{eq:u5}
  C_{\MP+\MA}(k;\alpha) = \sum_{j=-N+1}^{N-1} (N-|j|) \sigma^2
  \eee^{-\alpha|k-j|}.
\end{equation}
\begin{Conjecture}
  The MA filtering of the 1st order Markov process generates nearly a 2nd order
  Markov process.
\end{Conjecture}
The parameter of the 2nd order Markov process approximating the combined
(auto-) covariance \eref{eq:u5} can be obtained using the LS regression fit,
i.e.,
\begin{equation}\label{eq:u9}
  \hat{\alpha} = \argmin_{\hat{\alpha}}\ \sum_k\ (C_{\MP+\MA}(k;\alpha) -
  C_{\MPP}(k;\hat{\alpha}))^2.
\end{equation}
Substituting \eref{eq:x9} and \eref{eq:u5} into \eref{eq:u9}, and letting the
first derivative to be equal to zero, the LS estimate $\hat{\alpha}$ must
satisfy the linear equation,
\begin{equation*}
  \sum_k \hat{\alpha}|k|+1+W_{-1}\!\left(-\frac{C_{\MP+\MA}(k;\alpha)}
    {\eee}\right) = 0
\end{equation*}
which can be readily solved for $\hat{\alpha}$, and $W_{-1}$ denotes the
Lambert function \cite{Abramowitz74}.

\begin{figure}[H]
  \centering
  \psfrag{xx}[][][0.8]{$N$}
  \psfrag{yy}[][][0.8]{$\mse(\nx)$}
  \psfrag{zz1}[][][0.8]{$\alpha=0.1$}
  \psfrag{zz2}[][][0.8]{$\alpha=0.5$}
  \psfrag{zz3}[][][0.8]{$\alpha=0.9$}
  \includegraphics[scale=0.45]{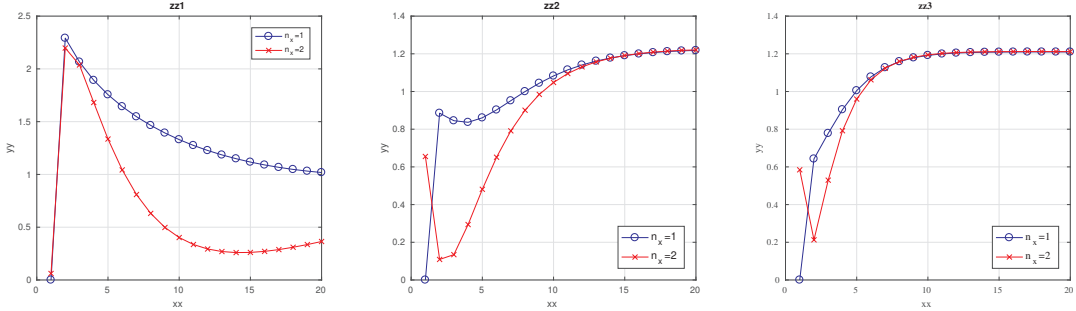}
  \caption{The MSE of approximating the (auto-) covariance of the 1st order
    Markov process at the output of length $N$ MA filter by the (auto-)
    covariance of the 2nd order Markov process.}\label{pict3}
\end{figure}

A discrete time sequence of $N$ elements has the (auto-) covariance constrained
to $(2N-1)$ time indexes as indicated in \eref{eq:u5}. Assuming the length $N$
MA filter, and that there are $(\nx N)$ samples, $\nx=1,2,\ldots$, of the 1st
order Markov process available, the (auto-) covariance \eref{eq:u5} has the
overall length $2N(\nx+1)-3$ samples. \fref{pict3} compares the MSE,
\begin{equation*}
  \mse(\nx) = 100 \times \frac{ \sum_k ( C_{\MP+\MA}(k;\alpha) -
    C_{\MPP}(k;\hat{\alpha}) )^2 }{ \sum_k C_{\MP+\MA}^2(k;\alpha) }  
\end{equation*}
of the LS fit of the (auto-) covariance of the 2nd order Markov process to the
combined (auto-) covariance of the 1st order Markov process and the impulse
response of the MA filter assuming three values of $\alpha$ and two values of
$\nx$. Given $\alpha$ and $\nx$, \fref{pict3} shows that the best LS fit occurs
for a certain value of the MA filter length $N$. It can be concluded that, in
general, the 1st order Markov process changes to the 2nd order Markov process
at the output of a MA filter.

The second problem to investigate is a linear MMSE (LMMSE) prediction of the
1st order Markov process observed at the output of a MA filter. In particular,
given $N$ samples $X_i=x(t_i)$, $i=1,2,\ldots,N$, of a random process having
the (auto-) covariance \eref{eq:u5}, the task is to predict its future value,
$X_{N+1}=x(t_{N+1})$, $t_{N+1}>t_N$. 

In general, the impulse response $\vh$ of the LMMSE filter to estimate an
unknown scalar parameter $P$ from measurements $\vX$ is computed as
\cite{Kay93},
\begin{equation}\label{eq:v1}
  \vh = \E{(\vx-\vxb)(\vX-\vXb)^T} \E{P\,\vX}.
\end{equation}
Here, the unknown parameter $P=X_{N+1}$, and
$\E{X_{N+1}X_i}= C_{\MP+\MA}(N+1-i)$ and $\E{X_iX_j} = C_{\MP+\MA}(i-j)$ in
\eref{eq:v1} which gives the LMMSE filter,
\begin{equation*}
  \vh = [ \underbrace{0\ 0\ \ldots\ 0}_{N-1}\ C_{\MP+\MA}(1)].
\end{equation*}
Consequently, the predicted value, $X_{N+1} = X_N\, C_{\MP+\MA}(1)$. Note that
the same procedure, but excluding the MA filter, gives the LMMSE estimate,
$X_{N+1} = X_N\, C_{\MP}(1)$.

The last problem to consider is a time alignment of two zero-mean, jointly
stationary processes. It is assumed that the normalized cross-covariance of
these two processes is,
\begin{equation}\label{eq:v3}
  \frac{\E{X_{1i}X_{2j}}}{\sqrt{\E{X_{1i}^2}\E{X_{2i}^2}}} =
  \eee^{-\alpha |i-j|} (1+\alpha|i-j|).
\end{equation}
Denote the uncertainty in determining the difference, $(i-j)$, as $\Delta$. In
order to estimate the unknown parameters $\alpha$ and $\Delta$, the left-hand
side of \eref{eq:v3} can be estimated by the method of moments, i.e., let
$\E{X_i^2} \approx \frac{1}{N} \sum_{i=1}^N X_i^2$, and,
$\E{X_{1i}X_{2j}} \approx \frac{1}{N} \sum_{i=1}^N X_{1i}X_{2|i-\Delta|}$. The
cross-covariance \eref{eq:v3} can be then rewritten as,
\begin{equation*}
  v_k = \frac{ \sum_{i=1}^N X_{1i}X_{2|i-\Delta-k|} }{\sqrt{\sum_{i=1}^N
      X_{1i}^2} \sqrt{\sum_{i=1}^N X_{2i}^2}} =  \eee^{-\alpha |\Delta+k|}
  (1+\alpha|\Delta+k|),\quad k=0,1,2,\ldots
\end{equation*}
Utilizing the Lambert function $W_{-1}$, the cross-covariance can be rewritten
further as,
\begin{equation}\label{eq:v7}
  \alpha|\Delta+k| = -1 - W_{-1}\!\left(-\frac{v_k}{\eee}\right) \triangleq
  \tv_k,\quad k=0,1,2,\ldots 
\end{equation}
Assuming, without loss of generality, that $\Delta\geq 0$, the absolute value
in \eref{eq:v7} can be ignored. Consequently, the unknown parameters $\alpha$
and $\Delta$ can be obtained as a linear LS fit to $N$ measured and calculated
values $\tv_k$ in linear model \eref{eq:v7}.

\section{Conclusions}

The development of a novel statistical measure to enable correlation analysis
for multiple random vectors resumed by summarizing background knowledge on
statistical description of discrete time random processes. This was then
extended with the derivation of several supporting results which were used in
the following sections. Specifically, it was shown that linear regression can
be effectively approximated by splitting the data into disjoint subsets, and
assuming only one average data point within each subset. In addition, a
procedure for generating multiple Gaussian processes with prescribed
auto-covariance and cross-covariance was devised. The main result of the paper
was obtained by assuming the Taylor's expansion of multivariate scalar
functions, and then approximating the Taylor's expansion by a univariate
polynomial. The single polynomial variable is a simple sum of variables in the
original multivariate function. The polynomial approximation represents a
mapping from multiple discrete time observations of a random process to a
scalar random field. The mean field value is a weighted sum of canonical
central moments with increasing orders. These moments were named central
sum-moments to reflect how they are defined. The sum-moments were then
discussed in light of other similar concepts such as total variance, Mean Field
Theory, and moving average sequence filtering. Illustrative examples were
studied in the last section of the paper. In particular, the accuracy of
approximate linear regression was evaluated quantitatively assuming two
disjoint data subsets. Assuming the 1st and the 2nd order Markov processes, the
central sum-moments were compared with the mean Minkowski distance. For
Gaussian processes, the central sum-moments can be obtained in a closed form.
The remaining problems investigated moving average filtering of the 1st order
Markov processes, and its prediction using a linear MMSE filter.

\vspace{6pt} 

\funding{This research received no external funding.}

\conflictsofinterest{The author declares no conflict of interest.}

\abbreviations{The following abbreviations are used in the paper:\\
  \noindent 
  \begin{tabular}{@{}ll}
    1MP & 1st order Markov process \\
    2MP & 2nd order Markov process \\
    AR & autoregressive \\
    LMMSE & linear minimum mean square error \\
    LS & least squares \\
    MMSE & minimum mean square error \\
    MSE & mean square error \\
    MA & moving average \\
    TV & total variance \\
  \end{tabular}\\

  \noindent The following symbols are used in the paper:\\
  \begin{tabular}{@{}cl}
    $|\cdot|$ & absolute value, set cardinality, sum of vector elements \\
    $\E{\cdot}$ & expectation\\
    $\corr{\cdot}$ & correlation\\
    $\cov{\cdot}$ & covariance\\
    $\var{\cdot}$ & variance\\
    $(\cdot)^{-1}$ & matrix inverse\\
    $(\cdot)^T$ & matrix/vector transpose\\    
    $f_x$ & distribution of a random variable $X$ \\    
    $f$, $\dot{f}$, $\ddot{f}$ & function $f$, and its first and second
                                 derivatives \\ 
    $\N_{+}$ & positive non-zero integers \\    
    $\R$, $\Rp$ & real numbers, positive real numbers\\
    $\Xb$ & mean value of random variable $X$ \\
    $X_{ij}$ & $j$-th sample of process $i$ \\
    $W_{-1}$ & Lambert function \\
  \end{tabular}
}


\reftitle{References}
\externalbibliography{yes}
\bibliography{refer}

\begin{thebibliography}{-------}
\providecommand{\natexlab}[1]{#1}

\bibitem[Drezner(1995)]{Drezner95}
Drezner, Z.
\newblock Multirelation - a correlation among more than two variables.
\newblock {\em Computational Statistics \& Data Analysis} {\bf 1995}, {\em
  19},~283--292.
\newblock
  doi:{\changeurlcolor{black}\href{https://doi.org/10.1016/0167-9473(93)E0046-7}{\detokenize{10.1016/0167-9473(93)E0046-7}}}.

\bibitem[Dear and Drezner(1997)]{Dear97}
Dear, R.; Drezner, Z.
\newblock On the significance level of the multirelation coefficient.
\newblock {\em Journal of Applied Mathematics \& Decision Sciences} {\bf 1997},
  {\em 1},~119--130.
\newblock
  doi:{\changeurlcolor{black}\href{https://doi.org/10.1155/S1173912697000114}{\detokenize{10.1155/S1173912697000114}}}.

\bibitem[Gei{\ss} and Einax(1996)]{Geis96}
Gei{\ss}, S.; Einax, J.
\newblock Multivariate correlation analysis - a method for the analysis of
  multidimensional time series in environmental studies.
\newblock {\em Chemometrics and Intelligent Laboratory Systems} {\bf 1996},
  {\em 32},~57--65.
\newblock
  doi:{\changeurlcolor{black}\href{https://doi.org/10.1016/0169-7439(95)00067-4}{\detokenize{10.1016/0169-7439(95)00067-4}}}.

\bibitem[Abdi(2007)]{Abdi07}
Abdi, H., Encyclopedia of Measurements and Statistics; SAGE,  2007; chapter
  Multiple correlation coefficient, pp. 648--655.
\newblock
  doi:{\changeurlcolor{black}\href{https://doi.org/10.4135/9781412952644}{\detokenize{10.4135/9781412952644}}}.

\bibitem[Sz\'ekely \em{et~al.}(2007)Sz\'ekely, Rizzo, and Bakirov]{Szekely07}
Sz\'ekely, G.J.; Rizzo, M.L.; Bakirov, N.K.
\newblock Measuring and testing dependence by correlation of distances.
\newblock {\em The Annals of Statistics} {\bf 2007}, {\em 35},~2769--2794.
\newblock
  doi:{\changeurlcolor{black}\href{https://doi.org/10.1214/009053607000000505}{\detokenize{10.1214/009053607000000505}}}.

\bibitem[Merig\'o and Casanovas(2011)]{Merigo11}
Merig\'o, J.M.; Casanovas, M.
\newblock A New Minkowski Distance Based on Induced Aggregation Operators.
\newblock {\em International Journal of Computational Intelligence Systems}
  {\bf 2011}, {\em 4},~123--133.
\newblock
  doi:{\changeurlcolor{black}\href{https://doi.org/10.1080/18756891.2011.9727769}{\detokenize{10.1080/18756891.2011.9727769}}}.

\bibitem[Nguyen \em{et~al.}(2014)Nguyen, M\"uller, Vreeken, Efros, and
  B\"ohm]{Nguyen14}
Nguyen, H.V.; M\"uller, E.; Vreeken, J.; Efros, P.; B\"ohm, K.
\newblock Multivariate maximal correlation analysis.
\newblock  International Conference on Machine Learning,  2014, pp.
  II--775--II--783.
\newblock
  doi:{\changeurlcolor{black}\href{https://doi.org/10.5555/3044805.3044979}{\detokenize{10.5555/3044805.3044979}}}.

\bibitem[Josse(2016)]{Josse16}
Josse, J.
\newblock Measuring multivariate association and beyond.
\newblock {\em Statistics Surveys} {\bf 2016}, {\em 10},~132--167.
\newblock
  doi:{\changeurlcolor{black}\href{https://doi.org/10.1214/16-SS116}{\detokenize{10.1214/16-SS116}}}.

\bibitem[Shu \em{et~al.}(2016)Shu, Zhang, Shi, and Qi]{Shu16}
Shu, X.; Zhang, Q.; Shi, J.; Qi, Y.
\newblock A comparative study on weighted central moment and its application in
  2D shape retrieval.
\newblock {\em Information} {\bf 2016}, {\em 7},~1--12.
\newblock
  doi:{\changeurlcolor{black}\href{https://doi.org/10.3390/info7010010}{\detokenize{10.3390/info7010010}}}.

\bibitem[B\"ottcher \em{et~al.}(2019)B\"ottcher, Keller-Ressel, and
  Schilling]{Bottcher19}
B\"ottcher, B.; Keller-Ressel, M.; Schilling, R.L.
\newblock Distance multivariance: New dependence measures for random vectors.
\newblock {\em The Annals of Statistics} {\bf 2019}, {\em 47},~2757--2789.
\newblock
  doi:{\changeurlcolor{black}\href{https://doi.org/10.1214/18-AOS1764}{\detokenize{10.1214/18-AOS1764}}}.

\bibitem[Wang and Zheng()]{Wang20}
Wang, J.; Zheng, N.
\newblock Measures of correlation for multiple variables.
\newblock \textbf{2020}, [math]arXiv:1401.4827.

\bibitem[Gardner(1990)]{Gardner90}
Gardner, W.A.
\newblock {\em Introduction to Random Processes With Applications}, 2nd ed.;
  McGraw-Hill,  1990.

\bibitem[Papoulis and Pillai(2002)]{Papoulis02}
Papoulis, A.; Pillai, S.U.
\newblock {\em Probability, Random Variables, and Stochastic Processes}, 4th
  ed.; McGraw-Hill,  2002.

\bibitem[Giller()]{Giller12}
Giller, G.L.
\newblock The Statistical Properties of Random Bitstreams and the Sampling
  Distribution of Cosine Similarity.
\newblock \textbf{2012}, SSRN preprint.,
  doi:{\changeurlcolor{black}\href{https://doi.org/10.2139/ssrn.2167044}{\detokenize{10.2139/ssrn.2167044}}}.

\bibitem[Kay(1993)]{Kay93}
Kay, S.M.
\newblock {\em Fundamentals of Statistical Signal Processing: Estimation
  Theory}; Vol.~I, Prentice Hall,  1993.

\bibitem[Oppenheim \em{et~al.}(2009)Oppenheim, Schafer, and Buck]{Oppenheim09}
Oppenheim, A.V.; Schafer, R.W.; Buck, J.R.
\newblock {\em Discrete-Time Signal Processing}, 3rd ed.; Prentice Hall,  2009.

\bibitem[Folland()]{Folland10}
Folland, G.B.
\newblock Higher-Order Derivatives and Taylor’s Formula in Several Variables.
\newblock \textbf{2010}, Math 425A Course notes.

\bibitem[Apostol(1973)]{Apostol73}
Apostol, T.
\newblock {\em Mathematical Analysis}, 2nd ed.; Pearson,  1973.

\bibitem[Yedidia(2001)]{Yedidia01}
Yedidia, J.S., Advanced Mean Field Methods; MIT Press,  2001; chapter An
  idiosyncratic journey beyond mean field theory, pp. 21--36.

\bibitem[Sobol(2001)]{Sobol01}
Sobol, I.M.
\newblock Global sensitivity indices for nonlinear mathematical models and
  their Monte Carlo estimates.
\newblock {\em Mathematics and Computers in Simulation} {\bf 2001}, {\em
  55},~271--280.

\bibitem[Saltelli \em{et~al.}(2005)Saltelli, Ratto, Tarantola, and
  Campolongo]{Saltelli05}
Saltelli, A.; Ratto, M.; Tarantola, S.; Campolongo, F.
\newblock Sensitivity analysis for chemical models.
\newblock {\em Chemical Reviews} {\bf 2005}, {\em 105},~2811--2828.
\newblock
  doi:{\changeurlcolor{black}\href{https://doi.org/10.1021/cr040659d}{\detokenize{10.1021/cr040659d}}}.

\bibitem[Shirali(2018)]{Shirali18}
Shirali, S.
\newblock {\em A Concise Introduction to Measure Theory}; Springer,  2018.

\bibitem[Boyd and Vandenberghe(2004)]{Boyd04}
Boyd, S.; Vandenberghe, L.
\newblock {\em Convex Optimization}; Cambridge University Press,  2004.

\bibitem[Abramowitz and Stegun(1974)]{Abramowitz74}
Abramowitz, M.; Stegun, I.A.
\newblock {\em Handbook of Mathematical Functions with Formulas, Graphs, and
  Mathematical Tables}; Dover,  1974.

\end{thebibliography}

\end{document}